\documentclass{article}
\usepackage[top=1.5in, bottom=1.5in, left=1.1in, right=1.1in]{geometry}
\pdfoutput=1
\usepackage{graphicx,epsfig,amssymb,amsmath, changepage,float, cite,appendix, epstopdf, bm, verbatim, caption,subcaption,amsfonts,dcolumn,dsfont,times,xcolor}
\usepackage[font=small,labelfont=bf]{caption}
\usepackage{pgfplots}
\usepackage{tikz}
\usetikzlibrary{positioning}
\usetikzlibrary{decorations.markings}
\usetikzlibrary{calc}

\usepackage{subcaption}

\numberwithin{equation}{section}
\numberwithin{figure}{section}
\numberwithin{table}{section}
\nocite{*}
\title{The homogenisation of Maxwell's equations with applications to photonic crystals and localised waveforms on metafilms}
\author{B. Maling, D.~J. Colquitt \& R.~V. Craster\\
Department of Mathematics, Imperial College London, London SW7 2AZ, U.K.}

\begin{document}

\maketitle





\section*{Abstract}\label{abstract}

An asymptotic theory is developed to generate equations that model the global behaviour of electromagnetic waves in periodic photonic structures when the wavelength is not necessarily long relative to the periodic cell dimensions; potentially highly-oscillatory short-scale detail is encapsulated through integrated quantities. 
Our approach is based on the method of high-frequency homogenisation (HFH), introduced in the context of scalar out-of-plane elastic waves in\cite{cras}, here extended to the full Maxwell system for electromagnetic waves in three dimensions. In doing so, a vector treatment of Maxwell's equations is necessary, of which scalar transverse electric (TE) and transverse magnetic (TM) polarised cases are a strict subset. The resulting procedure yields effective dynamic continuum equations, valid even at high frequencies, that govern a scalar function providing long-scale modulation of short-scale Bloch eigensolutions inside the structure. The form of the effective equation changes in the case of non-trivially repeated eigenvalues, leading to degeneracies that are linked to the appearance of Dirac-like points, a case that we explore using the asymptotic method. We examine the low-frequency long-wave limit in the case of dielectric structures, and find the expected result of classical homogenisation theory is captured, in which the usual governing equation is recovered, but now with the permittivity replaced by an effective tensor.

The theory we develop is then applied to two topical examples, the first being the case of aligned dielectric cylinders, which has great importance in the modelling of photonic crystal fibres. Advantage can be taken of the axial invariance to reduce the three-dimensional example such that it permits manageable computations on a two-dimensional domain. Results of the asymptotic theory are verified against these numerical simulations by comparing photonic band diagrams and evanescent decay rates for localised modes. We then consider the propagation of waves in a structured metafilm, here chosen to be a planar array of dielectric spheres. These waves can be highly localised in the plane, decaying exponentially in directions normal to the array, and at certain frequencies strongly directional dynamic anisotropy is observed. Computationally this is a challenging three-dimensional calculation, which we do here, and then demonstrate that the asymptotic theory captures the effect, giving highly accurate qualitative and quantitative comparisons as well as providing interpretation for the underlying change from elliptic to hyperbolic behaviour. 


\section{Introduction}\label{sec:intro}

The practical implementation of photonic crystal fibres (PCFs) in communications, lasers and other areas of engineering has resulted in much interest being ascribed to the dispersive properties of periodic photonic structures \cite{joannopoulos08a,zolla05a}, in particular those exhibiting photonic band gaps in one or more directions. Further to this, the rich band structures possessed by photonic crystals (PCs) give rise to a host of novel physical phenomena including, but not limited to, dynamic anisotropy \cite{chigrin03a}, ultrarefraction\cite{dirichlet} and light confinement\cite{confine} that can also impact upon the design of PCFs \cite{russell}. From a numerical point of view, the implementation of Maxwell's equations in such structures can pose considerable difficulty, particularly when the wavelength is of a similar scale to that of the repeating microcell, of which there may be many hundreds or more. In such cases, multiple scattering and interference play a central role. Typical numerical schemes such as finite element methods \cite{zolla05a} are highly effective, but the computational cost in tackling the coupled vector system increases rapidly with the number of nodes, which is intimately connected with the dimensionality, scale and operating wavelength of the system. Optimisation and design of a PC or PCF may require multiple simulations across changes in geometry, material parameters and microstructure, so it is attractive to explore complementary effective medium, or homogenisation, schemes \cite{belov05a,alu11a} that attempt to treat the medium in some averaged sense and use these instead of, or alongside, direct numerical simulation. 

Asymptotic homogenisation dates back to the French, Italian and Russian schools, and was developed for studying Partial Differential Equations (PDEs) with rapidly oscillating coefficients. The approach of introducing independent fast and slow variables applies naturally to structured media, leading to `effective medium' models on which there is extensive literature \cite{low1,low2,low4,bensoussan}. The effective properties attained using such methodologies are valid within the low-frequency (quasi-static) regime, which is the relevant domain for the rapidly developing theory of sub-wavelength metamaterials, in which a typical application involves tuning these effective properties to desired values as prescribed by transformation optics (see, for example \cite{cummer,dupont_wave}). Unfortunately these homogenisation theories often perform poorly near band edges \cite{belov05a} and the limitation of homogenisation theories to long wavelength excitations \cite{alu11a} means that most of the interesting effects associated with PCs, which are dynamic in nature and therefore correspond to high-frequency bands, are not captured.  

The article~\cite{cras} described how, by considering perturbations about the standing wave eigenfrequencies of a periodic structure, a long-scale PDE can be derived that governs the propagation of a scalar modulation function, characterising the group behaviour of waves in the structure. Crucially, this approach allows for arbitrarily rapid field variation inside the long-scale envelope, which in turn allows us to break free of the quasi-static regime, relying only on the conditions of Bloch's theorem\cite{bloch} being satisfied. Accordingly, we take advantage of a result from solid state physics:  the dispersive properties of a bulk periodic structure are  entirely characterised by the frequency dependence of the Bloch wave vector within the irreducible Brillouin zone\cite{kittel}. This result holds for general Bravais lattices, and recent work has been done to extend the current theory to such structures\cite{mehul14}. Here, we focus on simple square and cubic lattice arrays for clarity. Since publication of the asymptotic theory in 2010, further work has been done in which its results have been compared against direct numerical calculations, demonstrating that it is capable of capturing dynamic effects unique to high-frequency bands, including all-angle negative refraction, dynamic anisotropy and cloaking near Dirac-like cones\cite{dirichlet}, as well as reproducing the value of the reflection coefficient in simple scattering problems\cite{lina}. This supports the claim that the resulting behaviour can be considered that of an effective medium, and justifies our reference to the theory as high-frequency homogenisation (HFH). Alongside analogous work on discrete lattices\cite{lattice}, HFH was originally developed for the planar Helmholtz equation, which models the frequency-domain response of out-of-plane elastic shear waves, TE or TM polarised electromagnetic waves, or pressure waves in a fluid. Further work has been done to extend the theory to in-plane elasticity, in which the resulting Lam\'e system poses a vector problem in two dimensions\cite{tryfon_elastic,colquitt_grating}; see also \cite{boutin14a} for further work in elasticity.


The main result here is to extend the theory to the three-dimensional vector formulation of Maxwell's equations. We shall see that HFH leads to an effective PDE for a scalar modulation function $f_0$, in which the local dispersive properties of the structure are captured by a frequency-dependent tensor $T$. We show that in the low-frequency limit, this leads naturally to the expected result of classical homogenisation theory, in which the governing vector equation is recovered with effective material parameters.
 The theory we present is very general in its scope and we choose to illustrate its use on two relevant examples, the first being that of guided wave propagation in PCFs, and the second being that of dynamic anisotropy upon a structured metafilm. 

\begin{figure}[htbp]
  \centering
  \setlength{\unitlength}{\textwidth} 
\includegraphics[width=0.9\linewidth]{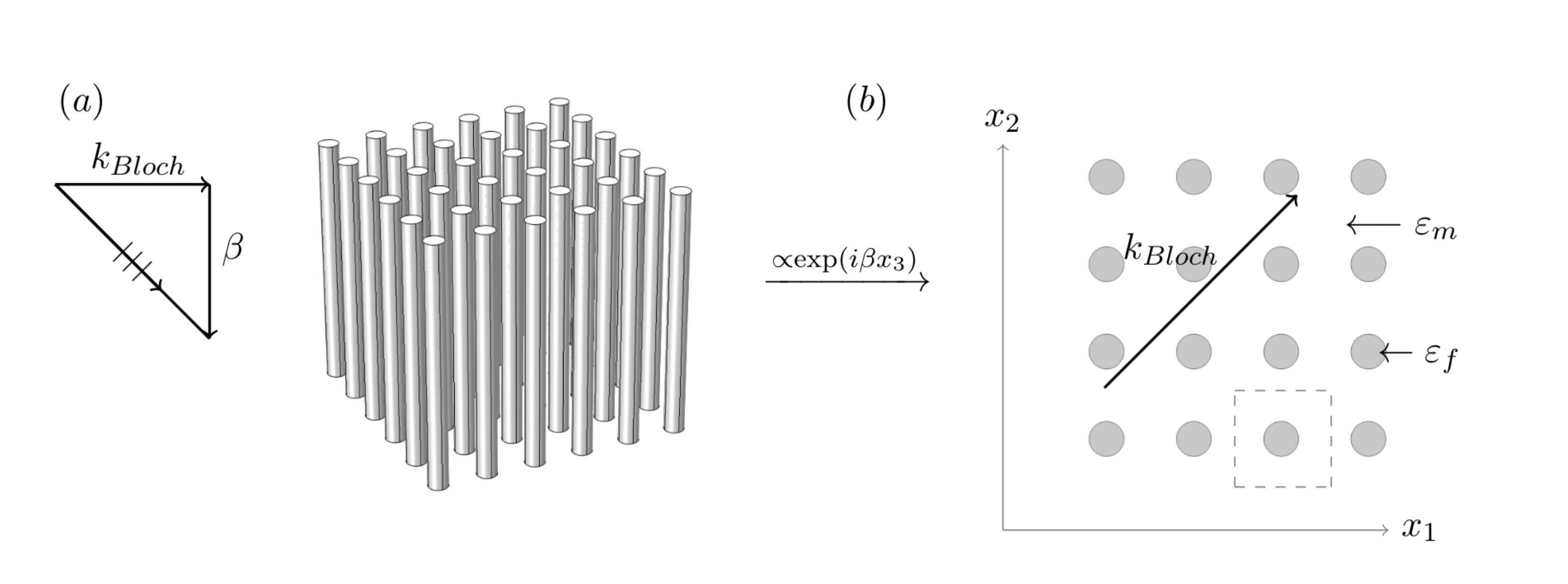}
\caption{{\scriptsize Reduction of a typical fibre-like structure to two dimensions. As shown in $(a)$, dielectric cylinders of permittivity $\epsilon_c$ are embedded in a homogeneous matrix phase with permittivity $\epsilon_m$. The resulting problem is projected onto the $(x_1,x_2)$ plane, and the angle of propagation depends on the ratio of the parameter $\beta$ and the planar Bloch wave vector, which characterises the phase shift across an irreducible cell (the dashed square in $(b)$)}}
\label{fig:reduce}
\end{figure}

The first application we consider is that of dielectric structures that are invariant in one spatial direction. The method could also be applied to so-called wire media \cite{lemoult11a}, but we note that dielectric structures are typical models for PCFs and, unlike metallic waveguides, support coupled modes at oblique incidence which are neither polarised in the transverse electric (TE) plane nor the transverse magnetic (TM) plane, and therefore demand a full vector treatment of Maxwell's equations. Compared with fully three-dimensional problems, this example has the advantage that the dependence of the field on the third spatial coordinate, say $x_3$, can be explicitly factored out as being proportional to  $\exp(i\beta x_3)$, reducing the problem to a quasi-two-dimensional one (see figure \ref{fig:reduce}), and therefore is a perfect example for testing the theory in a regime where full numerical simulations are tractable. Similar examples have been studied by various authors using numerical simulations\cite{guenneau_finite_1,guenneau_finite_2} as well as semi-analytic techniques such as multipole expansions\cite{guenneau_multipole}, and our current work complements these, adding further quantitative analysis and providing insight via the appealing notion of asymptotic homogenisation.

The final application we choose is that of  guided waves along a structured metafilm of dielectric spheres. Such structures \cite{holloway12a}, and metasurfaces \cite{maradudin11a,holloway12b},  are of considerable interest as a subclass of electromagnetic metamaterials that have advantages in terms of fabrication. We shall consider guided waves within a planar array of spheres that decay exponentially normal to the plane and, as such, are described as Rayleigh-Bloch waves \cite{wilcox84a,porter99a,colquitt_grating}.
In terms of the three-dimensional Maxwell system, Rayleigh-Bloch waves guided along a linear array of spheres was examined in~\cite{linton}.
Due to the computational effort required there have been very limited studies of the waves that can exist within planar array of spherical inclusions, with \cite{Thompson2010} focussing on scalar acoustics only, and multipole methods being used by \cite{shore07a}. See also \cite{holloway12a}; Rayleigh-Bloch waves are identified but only in limited directions. Here we access the full Brillouin zone numerically and identify frequencies for which marked dynamic anisotropy occurs, where the effective medium becomes locally hyperbolic, and as a result energy is localised and directed along characteristic directions. In simpler two-dimensional situations 
 star-shaped highly
directional wave motion at specific frequencies for structured media has emerged
in experiments and theory in optics \cite{chigrin03a,dangers}, and
is most strikingly seen in discrete mass-spring lattice systems 
\cite{slepyan08a,langley97b,dangers,colquitt2011}. Here for the full vector Maxwell system 
 the three-dimensional simulations are substantial and require many hours of computer time, whereas the asymptotic technique identifies these features, provides physical interpretation and finds quantitatively comparable results rapidly. 

The article is structured as follows: we begin by setting the scene in terms of the governing equations and necessary details regarding periodicity through reciprocal lattices, Brillouin zones and Bloch waves in section \ref{sec:three}. Given these preliminaries we then move on to developing the asymptotic theory in section \ref{sec:proc}, and use this as an opportunity to clarify some details regarding Dirac-like points (section \ref{sec:dirac}).
Furthermore, we confirm that the general theory provided does indeed retrieve the quasi-static classical homogenised result (section \ref{sec:low}). Applications to PCFs and then to dynamically anisotropic Rayleigh-Bloch waves on a metafilm  are given in sections \ref{sec:PCF}, \ref{sec:RB}, in which we use the asymptotic theory to complement full numerical simulations. Finally, some concluding remarks are drawn together in section \ref{sec:conclude}. 

\section{Governing equations and reciprocal lattice}\label{sec:three}


For time harmonic excitations proportional to $\exp(-i\omega t)$, and in the absence of sources, Maxwell's equations in linear media are given by \cite{jackson}:

\begin{equation}\begin{split}\label{maxwell}
\nabla\cdot\mathbf{D}&=0, \hspace{1.6cm}\nabla\times\mathbf{E}-i\omega\mathbf{B}=0,\\
\nabla\cdot\mathbf{B}&=0,\hspace{1.6cm}\nabla\times\mathbf{H}+i\omega\mathbf{D}=0,
\end{split}\end{equation}  

where $\bf E$ and $\bf B$ are the electric and magnetic fields respectively and are related to the displacement field $\bf D$ and magnetizing field $\bf H$ through the constitutive relations

\begin{equation}\label{const}
\mathbf{D}=\epsilon\mathbf{E}, \hspace{1.6cm}\mathbf{B}=\mu\mathbf{H}.
\end{equation}  

The parameters $\epsilon=\epsilon_{\text{r}}\epsilon_0$ and $\mu=\mu_{\text{r}}\mu_0$ are the permittivity and permeability respectively. 
 Our convention is to assume that the fields are complex vectors in $\mathbb{C}^3$ and the measurable fields are obtained by taking the real part. We assume that $\epsilon$ and $\mu$ are piecewise smooth, spatially varying real parameters that are exactly periodic in each direction. Combining the curl equations in (\ref{maxwell}) with the constitutive relations (\ref{const}) yields a decoupled equation satisfied by the magnetizing field:

\begin{equation}\label{max2}
-\nabla\times(\epsilon^{-1}\nabla\times\mathbf{H})+\mu\omega^2\mathbf{H}=0,
\end{equation}  

along with the condition that the component of both $\mathbf{E}$ and $\mathbf{H}$ parallel to a discontinuity in $\epsilon$ or $\mu$ must be continuous. In practice, both $\epsilon$ and $\mu$ are usually piecewise constant, and (\ref{max2}) has components that are Helmholtz equations coupled through boundary conditions between adjacent phases. We note that the magnetic formulation (\ref{max2}) turns out to be more convenient than the symmetric equation for $\mathbf{E}$ in the case of dielectrics, because the assumption that $\mu$ is constant ensures that $\nabla\cdot\mathbf{H}$ is identically zero, which is useful for the elimination of spurious modes that may arise in the numerical scheme\cite{guenneau_finite_1}. Perfect conductors can also be included in the structure, imposing a vanishing tangential electric field component at the boundary, corresponding to a vanishing curl of the magnetic field parallel to the boundary. This notion is meaningful only in the microwave region of the spectrum, which is significantly lower than the plasma frequency of many metals; such cases are of physical interest as this is the relevant regime for telecommunications. 

\begin{figure}[htbp]
  \centering
  \setlength{\unitlength}{\textwidth} 
\includegraphics[width=0.7\linewidth]{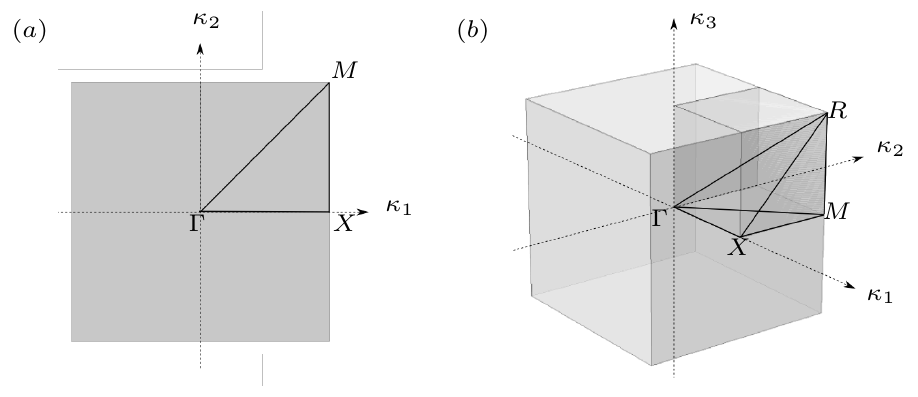}
\caption{{\scriptsize First Brillouin zones for $(a)$ simple square and $(b)$ simple cubic arrays. The triangle in (a), or  tetrahedron of (b), whose vertices are labelled are the \textit{irreducible} Brillouin zones, defined as the first Brillouin zones reduced by the point symmetries of the cell, in this case that of the square, cube respectively. Standing waves occur at each of the lettered points.}}
\label{fig:brillouin}
\end{figure}

The underlying periodicity of the medium that we seek to homogenise is an important ingredient of the asymptotic theory, and has a bearing on the numerical simulations we perform, and some notation is necessary which we provide here; this is available in standard texts \cite{brill2,kittel} amongst others. 
Suppose we have a Cartesian co-ordinate system with basis vectors ${\mathbf{\hat{x}_i}}$ for $i=1,2,3$. We assume that the periodicity of the structure is that of a simple cubic array, given explicitly by  $\epsilon (\mathbf{x})=\epsilon(\mathbf{x}+2l[m\mathbf{\hat{x}_1}+n\mathbf{\hat{x}_2}+p\mathbf{\hat{x}_3}])$ for $m,n,p\in\mathbb{Z}$, and similarly for $\mu(\mathbf{x})$. The primitive lattice vectors $2l\{\mathbf{\hat{x}_i}\}$ form an orthonormal set, and the elementary cell $\mathcal{C}$ can be chosen as any cube of side $2l$ oriented in accordance with these. Also central to our analysis will be the first Brillouin zone (shown in figure \ref{fig:brillouin} for a simple cubic lattice, along with the analogue for a square lattice in two dimensions), defined in terms of the reciprocal lattice vectors $\mathbf{k_i}$ for $i=1,2,3$, where $\mathbf{k_i}\cdot 2l\{\mathbf{\hat{x}_j}\}=2\pi\delta_{ij}$\cite{brillouin}. The relevance of reciprocal space here stems from Bloch's theorem, which states that for an infinite structure with direct lattice vectors $\{\mathbf{d}\}$, the electric field components are quasi-periodic such that $\mathbf{H}(\mathbf{x}+\mathbf{d})=\mathbf{H}(\mathbf{x})\exp(i\mathbf{k}\cdot\mathbf{d})$, and similarly for $\mathbf{E}$, where in this context the reciprocal vector $\mathbf{k}$ is called the Bloch wave vector\cite{bloch}. In particular, if the Bloch wave vector corresponds to a reciprocal lattice vector, the field has the periodicity of the direct lattice, resulting in a standing wave solution. In fact, standing wave solutions are found whenever the Bloch wave vector implies periodicity or anti-periodicity in each spatial direction, and it is these points about which the following asymptotic perturbation scheme is applied.

 \section{The asymptotic procedure}\label{sec:proc}

To adopt a two-scale approach we define the micro-scale position vector $\bm{\xi}=\mathbf{x}/l$, which is restricted such that inside the elementary cell $\mathcal{C}$ it takes values $\xi_i\in[-1,1]$ for $i=1,2,3$. Clearly, variation in $\epsilon$ and $\mu$ are functions of these co-ordinates only. From Bloch theory, we know that the three-dimensional structure admits standing wave solutions at discrete eigenfrequencies\cite{kittel}, corresponding to solutions with Bloch wave vectors at the vertices of the irreducible Brillouin zone. Perturbing the frequency slightly results in long wavelength modulation of the Bloch modes, and with this in mind we introduce a macro-scale position vector $\mathbf{X}=\eta\mathbf{x}/l$ relative to some fixed origin, with $0<\eta\ll 1$. Provided that the frequency perturbation is not too great, there is a natural separation of scales, with $\mathbf{X}$ representing the long-scale response of the structure. In the following theory, we are interested in the vanishing limit of $\eta$.

\begin{figure}[H]
  \centering
  \setlength{\unitlength}{\textwidth} 
\includegraphics[width=.7\linewidth]{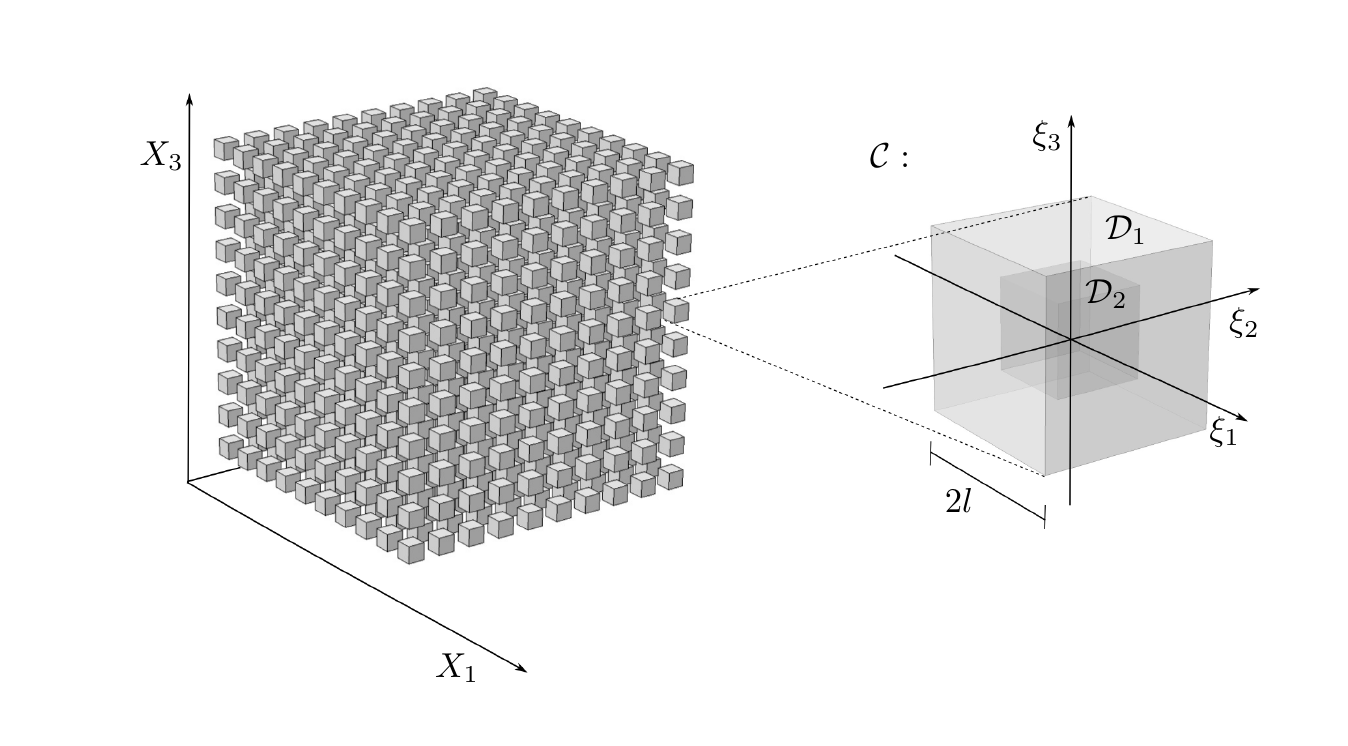}
\caption{{\scriptsize Disparate co-ordinate systems in a two-phase triply-periodic structure. The micro-scale co-ordinates ($\xi_1$,$\xi_2$,$\xi_3$) capture the field variation over a single elementary cell $\mathcal{C}=\mathcal{D}_1\cup\mathcal{D}_2$, whilst the macro-scale co-ordinates ($X_1$,$X_2$,$X_3$) encode the long-scale behaviour of the medium.}}
\label{fig:twoscale}
\end{figure}

 As is typical in homogenisation theories, the disparity of the length scales associated with $\bm{\xi}$ and $\mathbf{X}$ allows us to treat them as independent variables, so the partial derivative operators are expanded using the chain rule as $\partial_{x_i}=(\partial_{\xi_i}+\eta\partial_{X_i})/l$. We label the dimensionless standing wave eigenfrequencies $\Omega_0=\omega_0l/c$, with $c=1/\sqrt{\mu_0\epsilon_0}$ the speed of light in vacuum, corresponding to solutions satisfying periodic and/or antiperiodic boundary conditions on the cell level, explicitly given by

\begin{equation}\label{bcs}
\mathbf{H}|_{\xi_j=-1}=\pm \mathbf{H}|_{\xi_j=+1},\hspace{.5cm}\epsilon_{\text{r}}^{-1}\nabla\times\mathbf{H}|_{\xi_j=-1}=\pm\epsilon_{\text{r}}^{-1}\nabla\times\mathbf{H}|_{\xi_j=+1}\hspace{.5cm}j=1,2,3,
\end{equation}

 where $\pm$ is chosen depending on the vertex of the irreducible Brillouin zone. Following \cite{cras}, we seek solutions in the form of an asymptotic ansatz using the small parameter $\eta$:

\begin{equation}\label{ansatz}
\mathbf{H}=\sum\limits_{i=0}^\infty\eta^i \mathbf{H}_i,
\hspace{1.6cm}\Omega^2=\sum\limits_{i=0}^\infty\eta^i \Omega_i^2,
\end{equation}

 which leads to a hierarchy of equations to be solved upwards from the lowest order in $\eta$, with periodic/antiperiodic boundary conditions to be applied on the short scale at each level. The leading order system poses an eigenvalue problem on the short scale only:

\begin{equation}\label{O1}\mathcal{O}(1):
-\nabla_{\xi}\times\epsilon_{\text{r}}^{-1}(\bm{\xi})\nabla_{\xi}\times\mathbf{H}_0(\mathbf{X},\bm{\xi})+\mu_{\text{r}}(\bm{\xi})\Omega_0^2\mathbf{H}_0(\mathbf{X},\bm{\xi})=0,
\end{equation}

subject to short-scale periodicity conditions resulting from the two-scale expansion of (\ref{bcs}), and the following continuity conditions at any phase interfaces:

\begin{equation}\label{BCO1}
\big[\epsilon_{\text{r}}^{-1}\nabla_{\xi}\times\mathbf{H}_0\big]_{\partial\mathcal{D}_{1,2}}\times\mathbf{n}=0,\hspace{1.4cm}\big[\mathbf{H}_0\big]_{\partial\mathcal{D}_{1,2}}\times\mathbf{n}=0,\\
\end{equation}

 where $[\cdot]$ denotes a jump discontinuity and $\mathbf{n}$ is normal to the interface $\partial\mathcal{D}_{1,2}$ between two adjacent phases. For certain simple cases, it may be possible to construct analytic or semi-analytic solutions to this problem (for example, using multipole expansions for spherical or cylindrical inclusions, analogously to the methods used in \cite{guenneau_multipole,linton}). More often, however, the problem is solved numerically using a finite element package, say\cite{comsol}, which amounts to minimising the residue of the following weak equation: 

\begin{equation}
\int \Bigl (\epsilon_{\text{r}}^{-1}\nabla_\xi\times \mathbf{H}_0 \cdot
\nabla_\xi\times \mathbf{H}_0'^* +\delta(\nabla_\xi\cdot \mu_{\text{r}}\mathbf{H}_0)(
\nabla_\xi\cdot \mu_{\text{r}}\mathbf{H}_0'^*) - \mu_{\text{r}}\Omega_0^2
\mathbf{H}_0 \cdot \mathbf{H}_0'^* \Bigr ) dV = 0 \; ,
\label{Comsol}
\end{equation}

 where $\mathbf{H}_0'$ are the weight functions, $\delta$ is a small positive constant and $\cdot^*$ denotes complex conjugation. Although for $\Omega_0\neq 0$ it is usually unnecessary to further impose the divergence-free condition (which follows from taking the divergence of (\ref{O1})), we follow the approach of \cite{guenneau_finite_1} and include the second weak term in (\ref{Comsol}) to penalise spurious solutions that lie on the lowest branch and do not automatically satisfy this condition. This approach alleviates the restriction on geometry or choice of materials. We note that, as we are studying the vanishing limit $\eta$, the asymptotic hierarchy is robust even for high-contrast media, a subject that is of great interest in acoustic metamaterials, and approached in this context by Zhikov and others (see \cite{zhikov,smith}). In practice, a higher contrast will limit the frequency range over which we can expect our results to be reliable. Returning to the current problem, the general solution to (\ref{O1}) 
 is a linear combination of independent vector solutions whose coefficients must be allowed to vary on the long scale. Explicitly $H_{0i}(\mathbf{X},\bm{\xi})=f_0^{(r)}(\mathbf{X}) h_{0i}^{(r)}(\bm{\xi},\Omega_0)$ for $i=1,2,3$, and we sum over $r=1,2,...,p$ for an eigenvalue with multiplicity $p$. Each term is the product of a long-scale scalar modulation function $f_0^{(r)}(\mathbf{X})$ with a vector short-scale $h_{0i}^{(r)}(\bm{\xi},\Omega_0)$ that is a known Bloch eigensolution of (\ref{O1}); the aim is to find PDEs that the long-scale functions satisfy. 
 Note that the separated-scale component of functions are distinguished by lower-case lettering; this convention is followed for the remainder of this paper. We now move to the next order in the hierarchy, resulting in an inhomogeneous version of the leading order equation:
 
\begin{equation}\label{Oe}\mathcal{O}(\eta):
-\nabla_{\xi}\times\epsilon_{\text{r}}^{-1}\nabla_{\xi}\times\mathbf{H}_1+\mu_{\text{r}}\Omega_0^2\mathbf{H}_1=\nabla_{\xi}\times\epsilon_{\text{r}}^{-1}\nabla_{X}\times\mathbf{H}_0+\epsilon_{\text{r}}^{-1}\nabla_{X}\times\nabla_{\xi}\times\mathbf{H}_0-\mu_{\text{r}}\Omega_1^2\mathbf{H}_0.
\end{equation}

 with periodicity conditions as before and continuity conditions given by

\begin{equation}\label{BCOe}
\big[\epsilon_{\text{r}}^{-1}\left(\nabla_{\xi}\times\mathbf{H}_1+\nabla_{X}\times\mathbf{H}_0\right)\big]_{\partial\mathcal{D}_{1,2}}\times\mathbf{n}=0,\hspace{1cm}\big[\mathbf{H}_1\big]_{\partial\mathcal{D}_{1,2}}\times\mathbf{n}=0,
\end{equation}

 at the phase interfaces. Before attempting to solve this, we derive a compatibility condition (see appendix \ref{sec:Oe}), whose implication depends on the nature of the eigenvalue in question. The result resembles a $p$-component Dirac equation in long-scale position space:

\begin{equation}\label{Peq}
P_j^{nr}\frac{\partial f_0^{(r)}}{\partial X_j}+\Omega_1^2Q^{nr}f_0^{(r)}=0,
\end{equation}

\begin{equation*}\text{where}\hspace{.3cm} P_j^{nr}=i\Omega_0\int_C\left\{\mathbf{e_0}^{(n)*}\times\mathbf{h}_0^{(r)}+\mathbf{e_0}^{(r)}\times\mathbf{h}_0^{(n)*}\right\}_j\mathrm{d}V,\hspace{0.25cm}Q^{nr}=\int_C \mu_{\text{r}}\mathbf{h}_0^{(n)*}\cdot\mathbf{h}_0^{(r)}\mathrm{d}V.
\end{equation*}

Since the operator in (\ref{O1}) is Hermitian, we can always choose the eigenvectors $\mathbf{h}_0^{(n)}$ to be orthonormal such that $Q^{nr}=\delta_{nr}$, and this can be achieved using the Gram-Schmidt procedure. For fixed $n$, $P_j^{nn}$ is proportional to the $j$-component of the complex Poynting flux $\mathbf{e_0}^{(n)}\times\mathbf{h}_0^{(n)*}$ through the cell, which is necessarily zero for standing wave solutions satisfying the boundary conditions (\ref{BCO1}). This observation is easily proved from the fact that with (anti-)periodic boundary conditions it is always possible to normalise the fields such that either $\mathbf{e_0}^{(n)}$ or $\mathbf{h}_0^{(n)}$ is entirely real and the other is imaginary. The value of $P_j^{nr}$ for $n\neq r$ depends on the nature of the degeneracy under consideration; the first and most frequent type encountered is an \textit{essential} degeneracy, which results from the invariance of the system (i.e. the elementary cell and boundary conditions) under a particular symmetry transformation. For cubic and square arrays, such degeneracies can be shown by symmetry to yield $P^{nr}_j=0$. The second and more interesting case is that of an \textit{accidental} degeneracy, which occurs if we continuously tune the parameters of the structure until two otherwise distinct eigenvalues occur at the same frequency. Such degeneracies do not result from a symmetry group of the cell, and in general $P_j^{nr}\neq 0$ for $n\neq r$. Such cases are dealt with in section \ref{sec:dirac}.

 We deduce from (\ref{Peq}) that for an eigenvalue with $P^{nr}_j=0$ for every $n$ and $r$ then  $\Omega_1=0$ for non-trivial solutions. This results in locally quadratic dispersion bands passing through essentially degenerate eigenvalues (which is not necessarily the case for other Bravais lattices, as noted in \cite{chan_zero}). With this in place, we are in a position to solve (\ref{Oe}). The solution consists of a complimentary part, which has the same short scale form as the leading order solution, and a particular solution modulated by first order partial derivatives of $f_0$. The most general form thus has components given by $H_{1i}(\mathbf{X},\bm{\xi})=f_1^{(r)}(\mathbf{X})h_{0i}^{(r)}(\bm{\xi},\Omega_0)+f_{0,X_j}^{(r)}(\mathbf{X})h_{1ij}^{(r)}(\bm{\xi},\Omega_0)$. Exploiting the linearity of the system and the independence of the disparate position variables, (\ref{Oe}) then reduces to $3p$ separate systems of 3 coupled equations corresponding to each permutation of $j,r$. These are solved numerically using finite elements. We next move onto the $\mathcal{O}(\eta^2)$ system:

\begin{equation}\begin{split}\label{Oee}\mathcal{O}(\eta^2):
-\nabla_{\xi}\times\epsilon_{\text{r}}^{-1}\nabla_{\xi}\times\mathbf{H}_2+\mu_{\text{r}}\Omega_0^2\mathbf{H}_2&=\nabla_{\xi}\times\epsilon_{\text{r}}^{-1}\nabla_{X}\times\mathbf{H}_1+\epsilon_{\text{r}}^{-1}\nabla_{X}\times\nabla_{\xi}\times\mathbf{H}_1\\&+\epsilon_{\text{r}}^{-1}\nabla_{X}\times\nabla_{X}\times\mathbf{H}_0-\mu_{\text{r}}\Omega_2^2\mathbf{H}_0,
\end{split}\end{equation}

 with periodicity conditions as before and phase boundary conditions:

\begin{equation}\label{BCOee}
\big[\epsilon_{\text{r}}^{-1}\left(\nabla_{\xi}\times\mathbf{H}_2+\nabla_{X}\times\mathbf{H}_1\right)\big]_{\partial\mathcal{D}_{1,2}}\times\mathbf{n}=0,\hspace{2cm}\big[\mathbf{H}_2\big]_{\partial\mathcal{D}_{1,2}}\times\mathbf{n}=0.
\end{equation}

\begin{figure}[ht]
  \centering
\includegraphics[width=\linewidth]{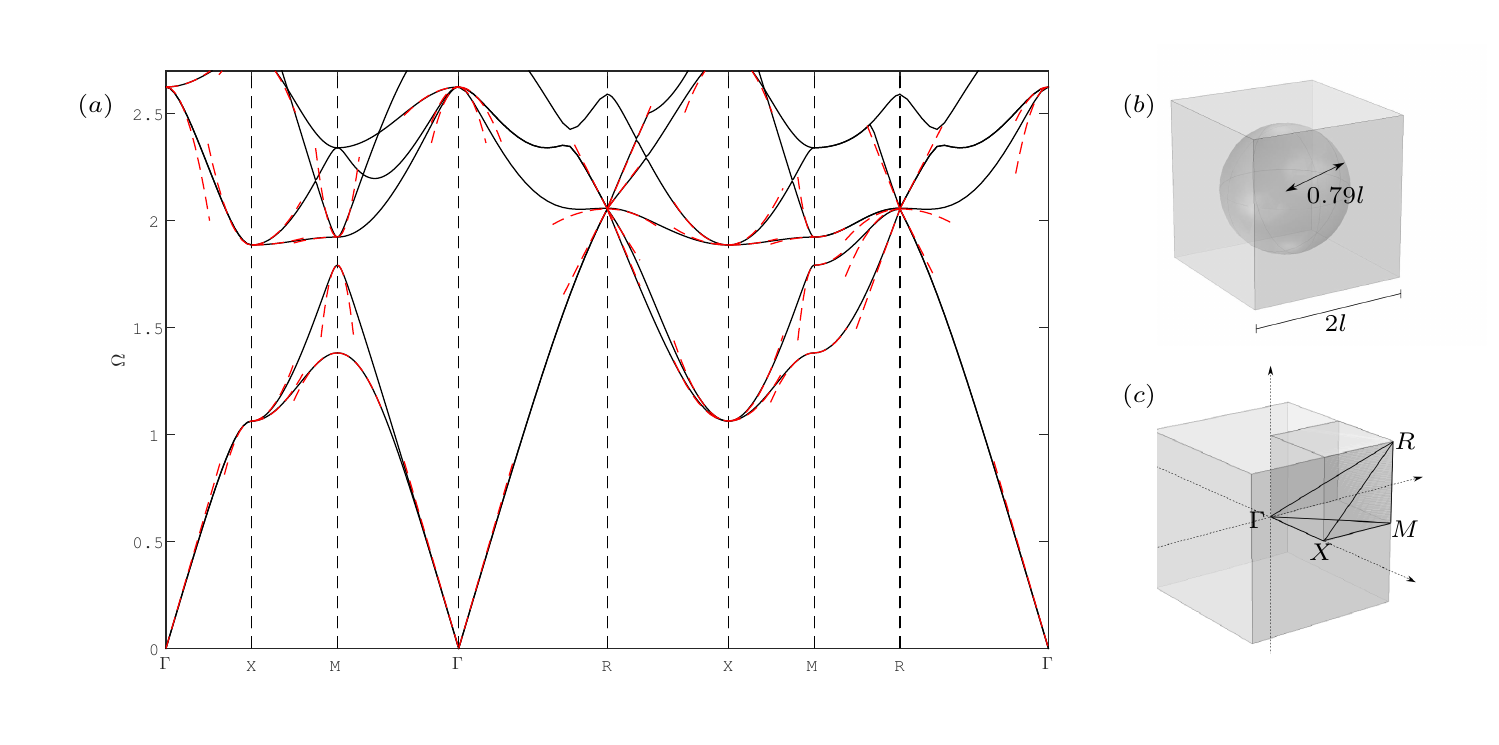}
\caption{{\scriptsize $(a)$ Photonic band diagram for a cubic array of perfectly-conducting spheres of radius $r=0.79l$ in a background of air, as shown in $(b)$. The band structure is plotted around the edges of the tetrahedral irreducible Brillouin zone $(c)$. Solid black curves are from FEM calculation, and red dashed curves are asymptotics from equation (\ref{pde8}). Note the Dirac-like point at $R$, which occurs only at this particular value of the radius.}}
\label{fig:balldisp}
\end{figure}

 Here we derive a second compatibility condition (see appendix \ref{sec:Oee}), analogous to that of the previous order. The result is an effective PDE satisfied by the modulation function $f_0(\mathbf{X})$, posed on the long-scale. For each independent eigenfunction at $\Omega_0$, $f_0$ is governed by

\begin{equation}\label{pde8}
T_{ij}\frac{\partial^2f_0}{\partial X_i \partial X_j}+\Omega_2^2f_0=0\Longleftrightarrow  T_{ij}\frac{\partial^2f_0}{\partial x_i \partial x_j}+\frac{(\omega^2-\omega_0^2)}{c^2}f_0=0,
\end{equation}

 where the components of the tensor $T_{ij}$ are formed from integrals over the cell involving the leading and first order solutions and are given in appendix \ref{sec:Oee}. The equation on the left makes clear the crucial point that the short scale is completely absent, having been integrated out and encapsulated by constant tensor $T_{ij}$, whereas the equation on the right has been rewritten in terms of dimensional variables, and makes clear that the the long-scale position variable $\mathbf{X}$ is an artefact of the asymptotic method. In a continuum setting, (\ref{pde8}) permits Bloch wave solutions of the form $f_0=\exp(i\bm{\kappa}\cdot \mathbf{X}/\eta)=\exp(i\bm{\kappa}\cdot \mathbf{x}/l)$, where the dimensionless quantity $\bm{\kappa}=(\mathbf{k}-\mathbf{k_0})l$ is proportional to the difference between the Bloch vector at frequency $\Omega$ and that at $\Omega_0$, corresponding to the vertex of the irreducible Brillouin zone. Substituting this into (\ref{pde8}) leads to locally quadratic behaviour of the dispersion bands as 
$\Omega_2^2=T_{ij}\frac{\kappa_i\kappa_j}{\eta^2}\implies\Omega=\Omega_0+T_{ij}\frac{\kappa_i\kappa_j}{2\Omega_0}$, where for the second relationship we have used the binomial theorem to expand the ansatz (\ref{ansatz}$(b)$) for small $\bm{\kappa}$. This serves as a useful verification of the asymptotic method, giving expressions for the local behaviour of the dispersion curves, as shown in figure \ref{fig:balldisp} for the case of perfectly-conducting spherical inclusions. The low-frequency limit, as well as the case of so-called Dirac-like points differ from this, both yielding locally linear dispersion and discussed in the following two sections.

\subsection{Dirac-like points}
\label{sec:dirac}

The unusual properties assciated with Dirac points in graphene \cite{neto09a} have prompted a surge of interest in the study of Dirac-like points in photonics \cite{chan_zero,dirichlet}. Such points are characterised by linear crossing of dispersion branches at Brillouin zone vertices, and in terms of the asymptotic theory are associated with non-trivial degeneracies of (\ref{O1}). In figure \ref{fig:balldisp}, a so-called accidental degeneracy is induced by choosing the radius of the spherical inclusion such that two eigenvalues coincide at $R$ point. In contrast, the other degeneracies observed in the band diagram are robust under the change of radius or material parameters, and can only be removed by breaking the symmetry of the cell; these are called essential degeneracies.

\begin{figure}[h]
  \centering
 \includegraphics[width=.7\linewidth]{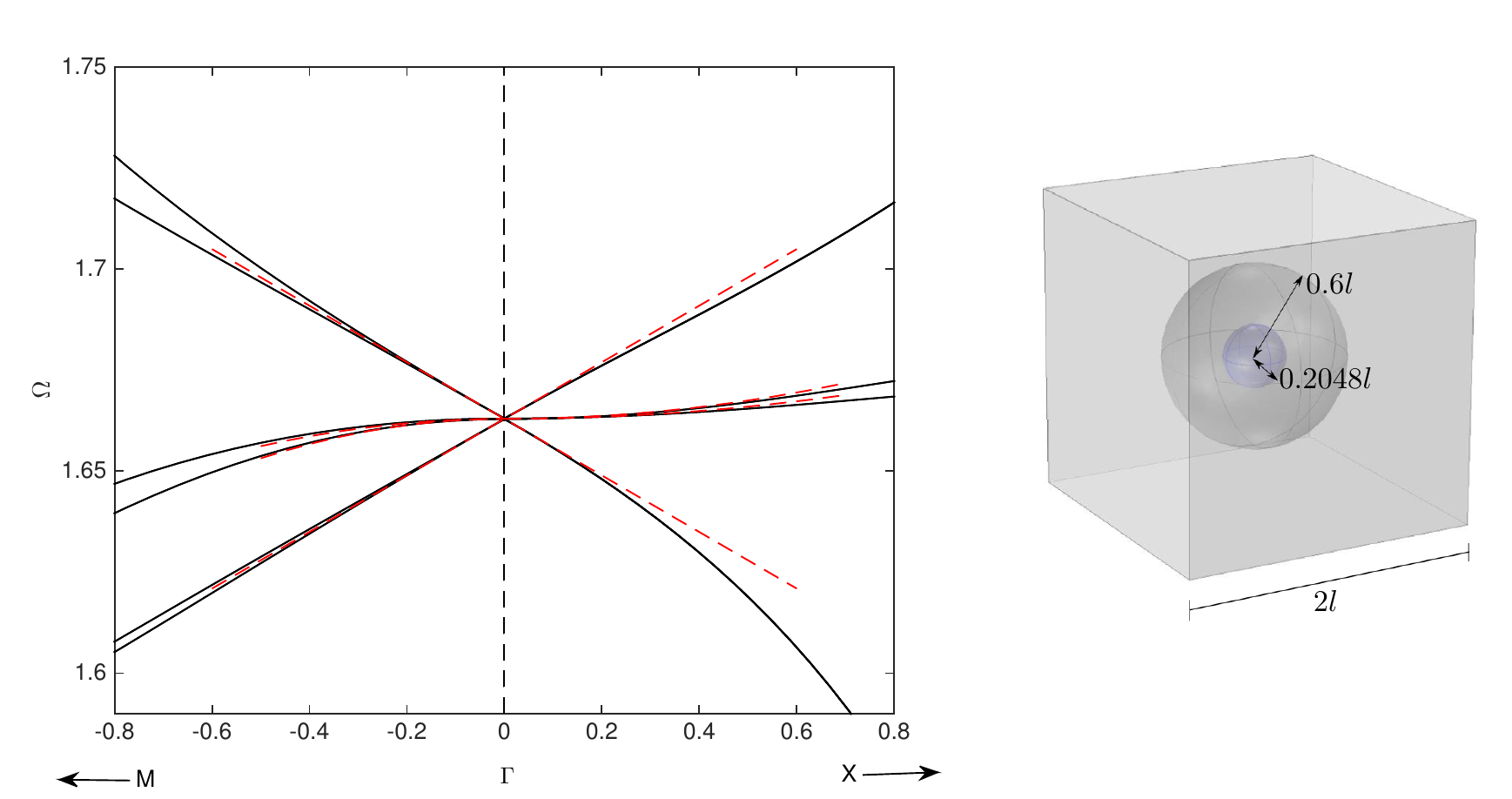}
\caption{{\scriptsize $(a)$ Dirac-like dispersion in a core-shell structure as demonstrated in \cite{chan_zero}. Here the air-filled elementary cell $(b)$ contains a spherical shell of permittivity 12 surrounding a perfectly conducting core. The effective equation governing the (repeated) linear branches is $0.054f_{0,X_1X_1}+0.054f_{0,X_2X_2}+0.054f_{0,X_3X_3}+\Omega_1^4f_0=0$, resulting from (\ref{Dtensor}).}}
\label{fig:dirac}
\end{figure}

 In the case of an accidental degeneracy (or incidentally for an essential degeneracy in certain non-cubic/square lattices), we find that the quantity $P_j^{nr}$ appearing in equation (\ref{Peq}) contains non-zero elements. As a result, the dispersion is governed to leading order by (\ref{Peq}), which results, unusually, in the appearance of locally linear dispersion at non-zero frequencies at a Brilouin zone vertex. It has been noted by various authors \cite{huang_zero,dirichlet}, that unusual wave guiding effects, as well as perfect transmission and cloaking can be observed in some instances around Dirac-like points. Such behaviour is considered in detail by Chan \textit{et al} in \cite{chan_zero}, where it is demonstrated using an effective medium theory\cite{wu_effective} that under certain conditions the dispersion can be mapped to that of a zero refractive index material. This is shown to occur if one can induce an accidental degeneracy between two triply degenerate eigenvalues at $\Gamma$ point, which is then demonstrated to be possible in a core-shell structure with perfectly conducting inclusions. The local dispersion in such cases is characterised by the crossing of two linear bands, with relatively flat quadratic bands passing through the middle, as seen in figure \ref{fig:dirac}. 

In order to extract the relevant asymptotics in the case of a Dirac-like point, we differentiate (\ref{Peq}) once with respect to $X_i$, and then by self-substitution obtain

\begin{equation}\label{Dtensor}
D_{ij}^{nr}\frac{\partial^2f_0^{(r)}}{\partial X_i \partial X_j}+\Omega_1^4f_0^{(n)}=0,
\end{equation}

 where $D_{ij}^{nr}=-C_j^{np}C_i^{pr}$ and $C_j^{nr}=(Q^{-1})^{np}P_j^{pr}$. Substituting the Bloch-wave solution $f_0^{(r)}(\mathbf{X})=\hat{f}_0^{(r)}\exp(i\bm{\kappa}\cdot \mathbf{X}/\eta)$ for fixed small $\bm{\kappa}$ leads to a homogeneous matrix equation that can be solved numerically to give the linear asymptotics, as has been done in figure \ref{fig:balldisp}, but in general it is not possible to obtain separate equations governing the evolution of the distinct modes. This is consistent with the assertion in \cite{chan_zero} that in general no effective medium description can be applied satisfactorily at Dirac-like points. Under certain circumstances, however, it does turn out to be possible; one such example being that of figure \ref{fig:dirac}, with (\ref{Dtensor}) decoupling to four identical isotropic equations: $0.054f_{0,X_1X_1}+0.054f_{0,X_2X_2}+0.054f_{0,X_3X_3}+\Omega_1^4f_0=0$, governing the linear bands, with a final pair yielding $\Omega_1=0$, and hence with quadratic dispersion governed by (\ref{pde8}). For the modes with linear dispersion, we deduce an effective Helmholtz equation for the Bloch wave, written in dimensional quantities as $\nabla^2 f_0+n_{\text{eff}}^2\epsilon_0\mu_0(\omega^2-\omega_0^2)f_0=0$, with $n_{\text{eff}}^2=\epsilon_0\mu_0(\omega^2-\omega_0^2)l^2/0.054\to 0$ as $\omega\to\omega_0$.

\subsection{The low-frequency long-wave limit in dielectric media}
\label{sec:low}

One surprising feature of HFH, when applied to a vector problem, is that the effective PDEs or systems of PDEs are scalar equations that do not mirror the form of the governing vector system of (\ref{max2}). This observation leads to an apparent inconsistency with classical homogenisation theory in the low-frequency regime, where the result is a vector equation resembling (\ref{max2}) but with effective tensors playing the role of $\epsilon$ and $\mu$\cite{guenneau_genhom}. In fact, using the current formulation, the effective permittivity and permeability are embedded in the system of equations (\ref{sys8}), which correctly describes the low-frequency linear asymptotics, but the two tensors cannot in general be extracted individually. However, in the commonly-occurring case of purely dielectric media, to which we now specify, the effective permeability $\mu_{\text{r}}\equiv 1$, and the effective permittivity tensor can then be extracted, as we now proceed to show.

Classical homogenisation theory is quasi-static and limited to dealing with situations where the harmonic frequency is sufficiently low that the field on the cell level is characterised by a small perturbation to an otherwise static field. This leads to dispersionless straight lines emanating from the $\Gamma$ point at zero frequency such as those seen in figure \ref{fig:balldisp}. Formally, this occurs when $\Omega^2=\mathcal{O}(\eta^2)$, corresponding to the substitution $\Omega_0=0$ into the leading order system, along with the divergence-free condition $\nabla_\xi\cdot\mathbf{H_0}=0$, which is no longer automatically satisfied by solutions of (\ref{O1}). Following the HFH procedure of section \ref{sec:proc}, we deduce that the leading order eigensolution is a constant vector field, and $\Omega_0=0$ is thus to be treated as a repeated eigenvalue with multiplicity $p=3$, corresponding to the three spatial directions. After some algebra, the resulting system of coupled effective PDEs can indeed be written as a single vector equation, which is the familiar result from classical homogenisation theory given in the context of dielectric media in \cite{guenneau_low}:

\begin{equation}\label{lowf}
-\nabla\times(\epsilon_{\text{r}}^{-1,\text{hom}}\nabla\times\mathbf{H}_0)+\omega^2\epsilon_0\mu_0\mathbf{H}_0=0,
\end{equation}

 where $\epsilon^{-1,\text{hom}}_{ij}=\langle \epsilon^{-1}\rangle \delta_{ij}+\tilde{\epsilon}^{-1}_{ij}$, and $\mathbf{H}_0$ is the leading order magnetic field. Here $\langle \cdot\rangle$ denotes a cell-averaged quantity, and the correction tensor $\tilde{\epsilon}^{-1}_{ij}$ has components depending on the geometry of the cell. To prove this, we note that in the quasi-static limit $\Omega_0=0$, the leading order solutions can be chosen without loss of generality to be constant vectors oriented along each of the three co-ordinate axes so that $h_{0i}^{(n)}=\delta_{ni}$, and hence $\mathbf{H}_0=(f_0^{(1)},f_0^{(2)},f_0^{(3)})^T$. We need to access the algebra of appendix \ref{sec:Oee} and compare equation (\ref{lowf}) to the HFH result for repeated eigenvalues, (\ref{sys8}): the two are identical if we can write $T^{nr}_{ij}=\varepsilon_{nik}\varepsilon_{jrl}\epsilon_{kl}^{-1,\text{hom}}$ for some second-rank tensor $\epsilon_{kl}^{-1,\text{hom}}$, which is equivalent to the statement that the tensor  $T$ contains the following symmetries: $T_{ij}^{nr}=T_{nr}^{ij}=-T_{ir}^{nj}=-T_{nj}^{ir}$.  Substituting the constant fields into (\ref{sys8}) we find the explicit form for $T$ as

\begin{equation}\label{Alow}\begin{split}
T_{ij}^{nr}=\int_\mathcal{C} \epsilon_{\text{r}}^{-1}\left(\delta_{ij}\delta_{nr}-\delta_{jn}\delta_{ir}+\partial_ih_{1nj}^{(r)}-\partial_nh_{1ij}^{(r)}\right)\mathrm{d}V.
\end{split}\end{equation}

 It is straightforward to show that the first two terms in the integrand satisfy the necessary symmetries, which are inherited directly from the $-\nabla_X\times\nabla_X\times(\cdot)$ operator in (\ref{pde1}). It is also trivial to note that the remaining two terms are antisymmetric in $n$ and $i$. The only non-trivial symmetry that remains is antisymmetry of the second two terms with respect to $j$ and $r$. In order to prove this, we consider the inhomogeneous problem (\ref{Oe}): substituting in the constant leading order solutions, along with the candidate particular solution $H_{1i}=f_{0,X_j}^{(r)}h_{1ij}^{(r)}$ leads to the following coupled problems for $i,j,r\in \{1,2,3\}$:

\begin{equation}\label{symoe}
\partial_k(\epsilon_{\text{r}}^{-1}\partial_kh_{1ij}^{(r)})-\partial_k(\epsilon_{\text{r}}^{-1}\partial_ih_{1kj}^{(r)})=\delta_{ij}\partial_r\epsilon_{\text{r}}^{-1}-\delta_{ir}\partial_j\epsilon_{\text{r}}^{-1},
\end{equation}

subject to boundary conditions at any phase interface given by

\begin{equation}\label{symoebc}
\left[\epsilon_{\text{r}}^{-1}\partial_kh_{1ij}^{(r)}n_k-\epsilon_{\text{r}}^{-1}\partial_ih_{1kj}^{(r)}n_k\right]_{\mathcal{D}_{1,2}}=\Big[\epsilon_{\text{r}}^{-1}\Big]_{\mathcal{D}_{1,2}}(\delta_{ij}n_r-\delta_{ir}n_j),
\end{equation}

where $n_i$ is the component of the unit normal $\mathbf{n}$ in the $x_i$-direction and $[\cdot]$ denotes a jump discontinuity. The two-scale expansion of the divergence-free condition $\nabla_\xi\cdot\mathbf{H_1}=-\nabla_X\cdot\mathbf{H_0}$ must also be imposed, which gives $f_{0,X_j}^{(r)}h_{1ij,\xi_i}^{(r)}=-f_{0,X_j}^{(r)}\delta_{jr}$. In the case that $j\neq r$, this condition is homogeneous, and because the right hand sides of both (\ref{symoe}) and (\ref{symoebc}) are antisymmetric in $j,r$, we deduce that $h_{1ij}^{(r)}=-h_{1ir}^{(j)}$. On the other hand, for $j=r$ the divergence-free condition yields $h_{1ij,\xi_i}^{(r)}=-1$. Integrating this over the cell $\mathcal{C}$ and applying the divergence theorem and periodicity leads to a contradiction $``0=1"$ that can only be avoided if $f_{0,X_j}^{(r)}\delta_{jr}=0$. We have hence derived the long-scale divergence-free condition $\nabla_X\cdot\mathbf{H_0}=0$, and with this in place we can show for $j=r$ that $h_{1,ij}^{(r)}=0$. In doing so, $T$ is deduced to contain the correct symmetries, and we derive an expression for the effective permittivity as follows:

\begin{equation}\label{epslow}
T^{nr}_{ij}=\varepsilon_{nik}\varepsilon_{jrl}\epsilon_{kl}^{-1,\text{hom}}\implies\epsilon_{kl}^{-1,\text{hom}}=\frac{1}{(2!)^2}\varepsilon_{nik}\varepsilon_{jrl}T^{nr}_{ij}.
\end{equation}

For each permutation of $k$, $l$, we only need evaluate one component of $T$. It is natural to split the resulting inverse permittivity tensor into a sum of two parts, the first resulting from the Kronecker delta terms in the integrand of $T_{ij}^{nr}$, given by $\langle\epsilon^{-1}\rangle\delta_{ij}$, and the second from the two remaining terms, which we refer to as the correction tensor $\tilde{\epsilon}_{ij}^{-1}$. Putting all this together we recover (\ref{lowf}) and hence the quasi-static homogenisation of Maxwell's equations.

\section{Application to PCFs}
\label{sec:PCF}

We now turn our attention to the case of dielectric cylinders of infinite length and constant permittivity $\epsilon_c$ aligned in the $x_3$-direction and embedded in a matrix phase with permittivity $\epsilon_m$. This model is appropriate for both holey-type PCFs, in which the cylinders are air holes in a dielectric background, as well as ARROW-type PCFs, in which the cylinders have a higher refractive index than the background. As discussed in section \ref{sec:intro}, the $x_3$-dependence of the solution can be factored out as $\propto\exp(i\beta x_3)$ where $\beta$ is the propagation constant of the radiation.

\begin{figure}[htbp]
  \centering
  \setlength{\unitlength}{\textwidth}
\includegraphics[width=\linewidth]{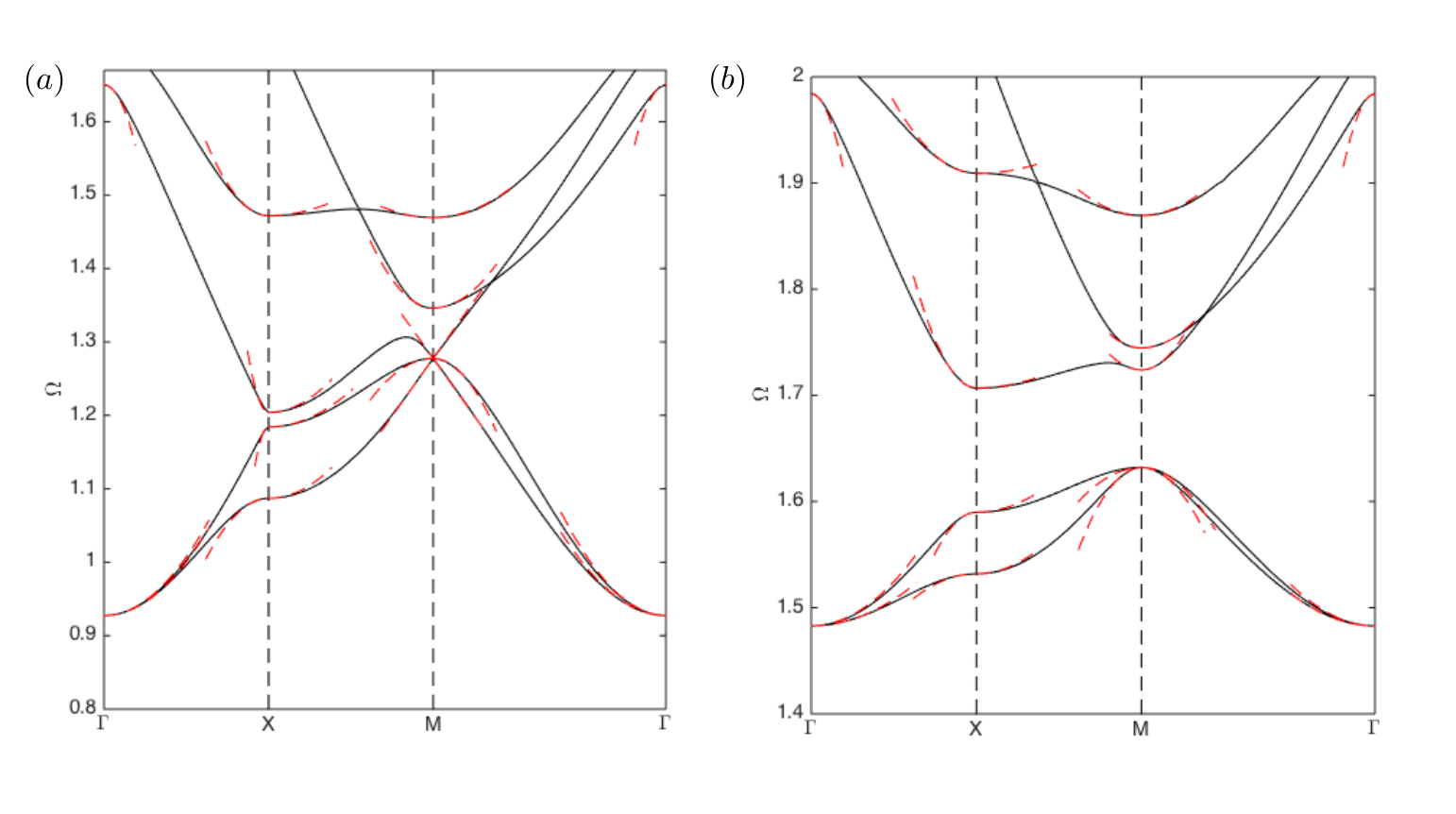}
\caption{{\scriptsize Transverse photonic band diagrams for the lowest 5 bands for a square array of cylindrical air holes of radius $r=0.75l$, where $2l$ is the pitch of the array, in a dielectric background with relative permittivity $\epsilon_{\text{r}}=6$. In $(a)$, the parameter $\beta=0.89/d$ has been tuned to induce an accidental degeneracy, resulting in a Dirac-like point at $M$. In $(b)$, $\beta$ has been increased to $3/l$, causing the higher-frequency bands to separate from the lowest two, so the PCF exhibits a partial band gap in the transverse plane. Note that different scales have been used on the vertical axes.}}
\label{fig:fibresdisp}
\end{figure}

For a fixed value of $\beta$, the resulting Bloch wave structure is projected onto the $(x_1,x_2)$ plane and the effective PDEs will govern propagation in these directions. Tuning $\beta$, along with the geometric and material properties of the cell, can then be used to induce particular features, such as partial stop-bands or Dirac-like points in the transverse plane. In figure \ref{fig:fibresdisp}, we show the band diagrams for one such configuration of air holes in a dielectric background at two different values of $\beta$ chosen such that a transverse stop-band and a Dirac-like point are exhibited. 
 
 In order to obtain the quasi-planar analogues of (\ref{Peq}) and (\ref{pde8}), we simply make the substitutions $\partial/\partial X_3\to0$ and $\partial/\partial\xi_3\to i\beta$, and the corresponding integrals become surface integrals over the two-dimensional cell. It is also straightforward to check for $\beta=0$ that our theory reduces to that of planar HFH, as published in \cite{cras,neumann}.

\subsection{Transverse stop bands and localised defect modes}

\begin{figure}[htbp]
  \centering
  \setlength{\unitlength}{\textwidth}
\includegraphics[width=.9\linewidth]{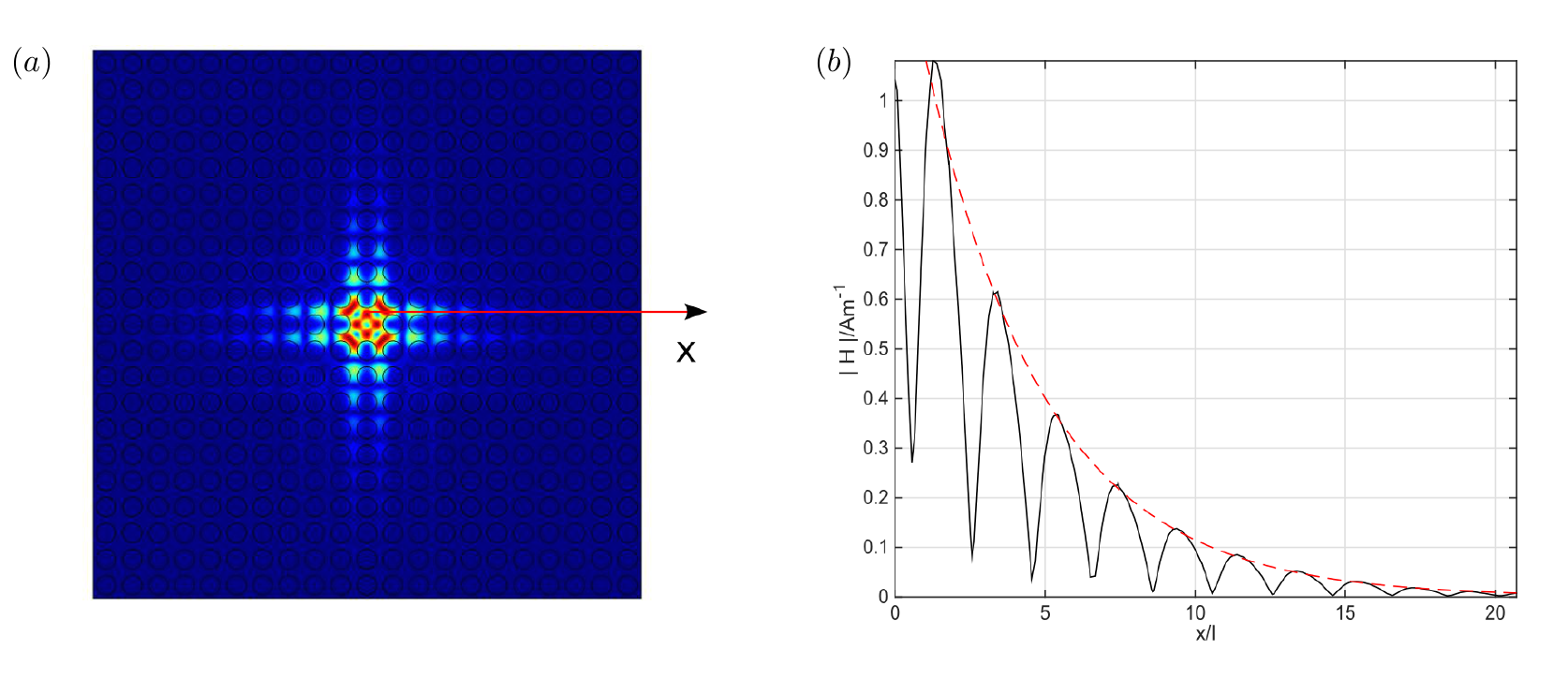}
\caption{{\scriptsize Magnetic field norm for a localised defect mode at $\Omega=1.690$, induced by removing one air cylinder from a large array of cells (of which a portion are shown here), corresponding to those of fig \ref{fig:fibresdisp}($b$). The dominant effective equation is $1.01f_{0,X_1X_1}+0.098f_{0,X_2X_2}+\Omega_2^2f_0=0$, where $\Omega_0=1.707$, corresponding to the third eigenvalue at $X$, which leads to directed leakage in the $x_1$-direction, and hence also in the $x_2$-direction due to the symmetry point at the opposite vertex of the first Brillouin zone. The decay of the mode in these directions is then very well approximated by an exponential form $\exp(-\alpha x)$ (the red dashed line in ($b$)), where the decay rate $\alpha=0.24/l$ follows from the PDE.}}
\label{fig:defect}\end{figure}

So far we have focused on the application of HFH to propagating `slow-modes' inside a bulk periodic structure, and the corresponding modes in the quasi-planar case propagate in the transverse plane as well as in the axial direction.  Alternatively, we may consider a structure exhibiting a transverse stop band at a particular value of $\beta$. If we introduce a finite defect of some kind into the structure, it may be possible to excite a mode whose frequency lies within the stop band, and hence decays evanescently in the surrounding medium. If the frequency lies in the vicinity of one of the eigenfrequencies $\Omega_0$, the exponentially decay is then governed by (\ref{pde8}). In figure \ref{fig:defect}, we show one such mode, where a defect is induced by removing one air cylinder from a large array of cells corresponding to those of fig \ref{fig:fibresdisp}($b$). Such a model is highly relevant for real-life applications of PCFs, and one can envisage the extension of such work giving rise to the notion of `effective fibres'. 

\section{Application to Rayleigh-Bloch modes}
\label{sec:RB}

For various wave systems, it is well known that periodic arrays of defects in otherwise homogeneous media can support localised modes.
Such modes exhibit Bloch wave quasi-periodicity in the array, and decay evanescently in the surrounding medium, reminiscent of Rayleigh waves at the surface of solids.
It is therefore natural to describe them as Rayleigh-Bloch modes, and their existence is ubiquitous in systems where the dimensionality of the array is at least one less than that of the wave system in question.
Rayleigh-Bloch waves are distinguished from ``pure'' surface waves (e.g. Rayleigh waves, Lamb waves, etc.) in that they can propagate along surfaces that exhibit some material or geometric periodicity and which, in the absence of this periodicity, do not support surface waves.
In this sense they are similar to spoof surface plasmons~\cite{Pendry2004}.
In two- and three-dimensional systems, linear arrays may be employed as diffraction gratings, and have been studied in the context of water waves and acoustics as solutions to the Helmholtz equation~\cite{Porter2005,Linton2007,Thompson2010}, surface plasmons~\cite{Maier2006,Navarro-Cia2009}, as well as coupled elastic waves~\cite{colquitt_grating}. In three-dimensional systems, such as the vector Maxwell system, Rayleigh-Bloch modes can exist for linear or planar arrays\cite{linton}, the latter allowing for the possibility of a range of in-plane effects to be observed.

\begin{figure}[h]
\centering
\includegraphics[width=0.45\linewidth]{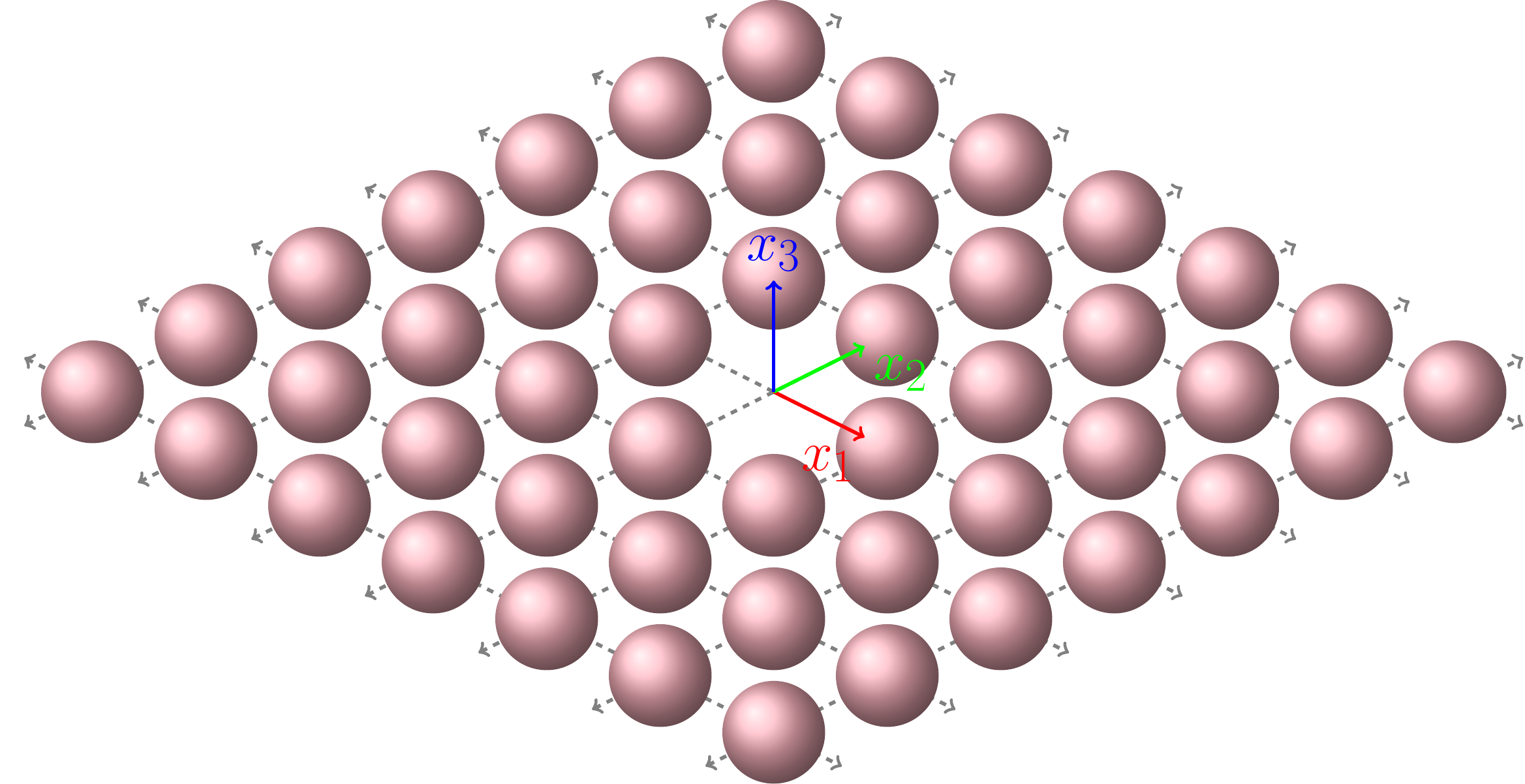}
\caption{\label{fig:RB-diagram}\scriptsize
A Doubly-periodic planar square array of dielectric spheres.
For the purpose of computation, the spheres have radii of $0.8l$, where the pitch of the array is $2l$.
The spheres are embedded in free space and have a relativity permittivity $\epsilon_{\text{r}}=20$.}
\end{figure}

\begin{figure}
  \centering
  \setlength{\unitlength}{\textwidth}
\includegraphics[width=.9\linewidth]{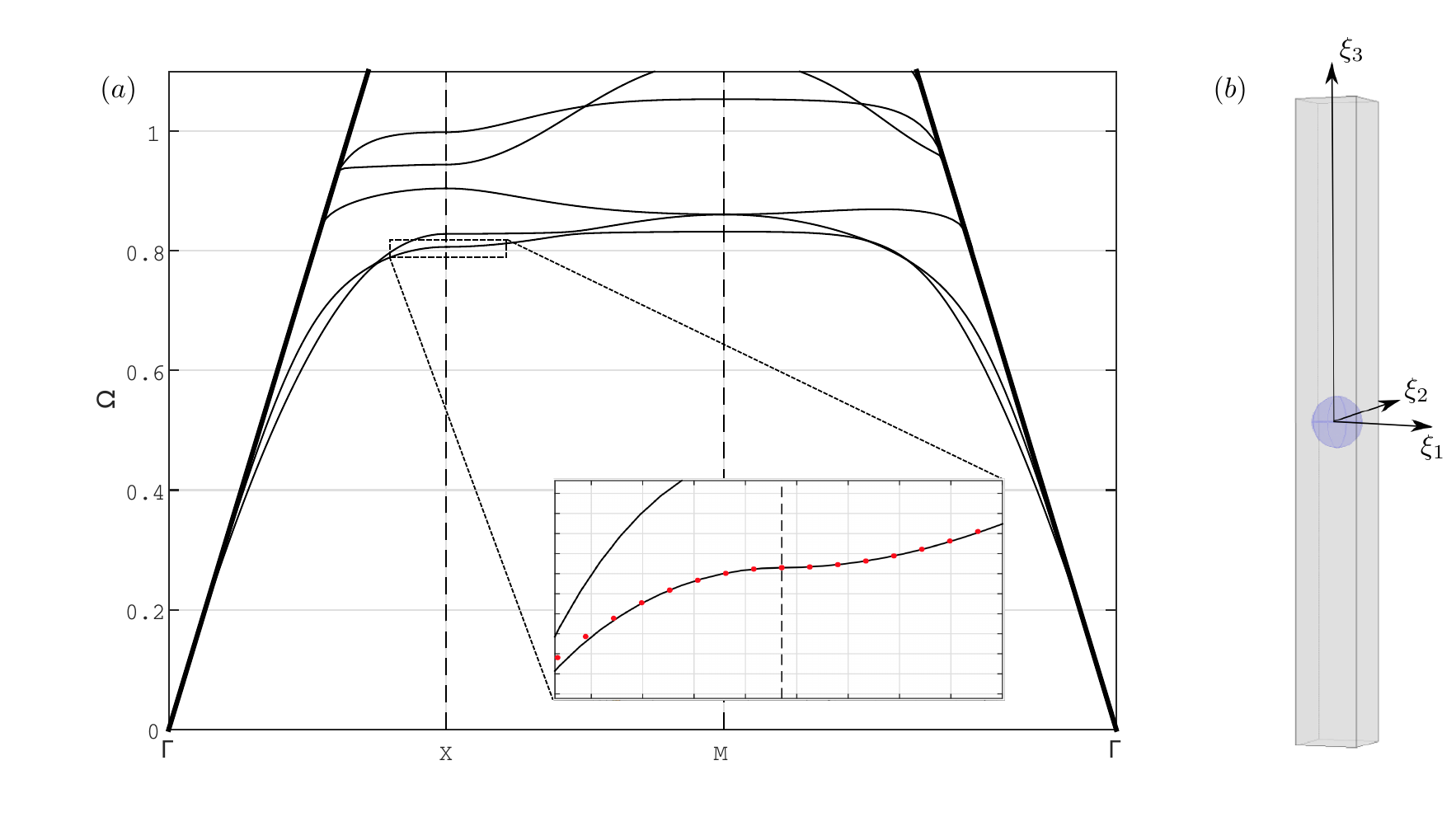}
\caption{{\scriptsize $(a)$ In-plane photonic band diagram for a square array of high-permittivity ($\epsilon_{\text{r}}=20$) dielectric spheres in vacuum, as shown in figure~\ref{fig:RB-diagram}. The radius of the spheres is $0.8l$, where $2l$ is the pitch of the array, corresponding to the elementary cell $(b)$. Here the cell is extended in the $\xi_3$-direction, and perfectly conducting boundary conditions are applied at its top and bottom of the cell. The inset of $(a)$ shows an enlarged portion of the diagram in which the local behaviour is governed by the hyperbolic effective equation $-0.19f_{0,X_1X_1}+0.11f_{0,X_2X_2}+\Omega_2^2f_0=0$, leading to the asymptotics shown by red dots.}}
\label{fig:RBdisp}
\end{figure}

The asymptotic technique we have developed is particularly well-suited to studying electromagnetic Rayleigh-Bloch systems, as the corresponding full numerical computations can be prohibitively demanding, particularly for fully coupled three-dimensional problems of the type examined here.
In order to assess and demonstrate the efficacy of the asymptotic technique developed in the preceding sections, we consider an infinite doubly-periodic planar array of dielectric spheres embedded in free space, as depicted in figure~\ref{fig:RB-diagram}.
Once again, the theory is largely unaltered from the three-dimensional case; we simply make the substitution $\partial/\partial X_3\to0$ in (\ref{Peq}) and (\ref{pde8}), and extend the cell in the $x_3$ direction as shown in figure \ref{fig:RBdisp}$(b)$. The corresponding photonic band diagram is shown in \ref{fig:RBdisp}$(a)$.

\subsection{Planar dynamic anisotropy \& hyperbolic metafilms}

The band diagram of figure~\ref{fig:RBdisp} and has several interesting features including a planar band-gap as well as flat bands associated with so-called \emph{slow waves}. A particularly fascinating feature is the change in curvature of the lowest branch at lattice point $\mathrm{X}$ in the reciprocal space, shown in the inset of figure~\ref{fig:RBdisp}$(a)$ along with the corresponding local asymptotics.
Such saddle points are associated with dynamic anisotropy~\cite{Osharovich2010,colquitt2011,Craster2012} and hyperbolic materials~\cite{Poddubny2013,Shekhar2014}; this type of media, when excited by an appropriate multipole source, direct the electromagnetic energy along clearly defined characteristic lines.
The existence of these saddle points for the system under consideration suggests the interesting possibility of dynamic anisotropy in a Rayleigh-Bloch wave setting, creating an effective hyperbolic metasurface.

\begin{figure}
\centering
\begin{subfigure}{0.45\linewidth}
\includegraphics[width=\linewidth]{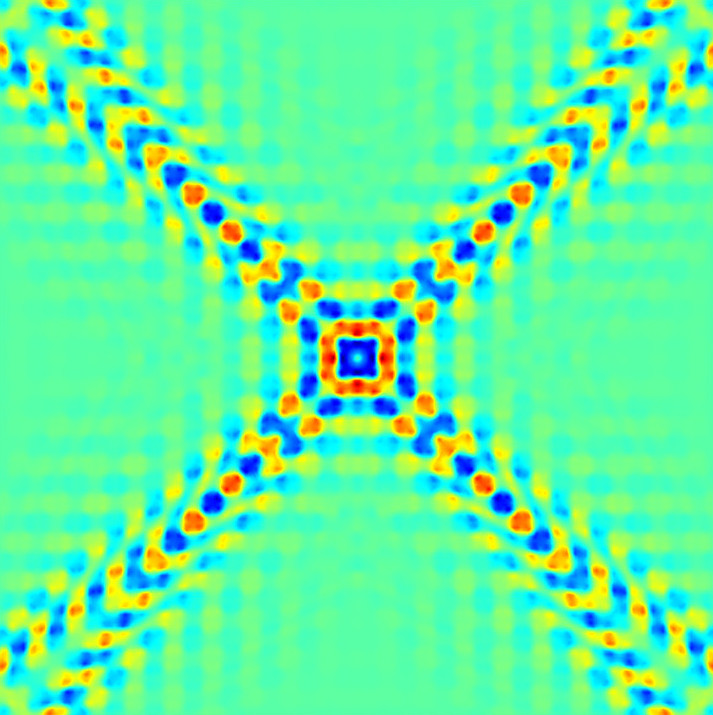}
\caption{\label{fig:RB-waves-HFH}\scriptsize
Asymptotic analysis}
\end{subfigure}
\begin{subfigure}{0.45\linewidth}
\includegraphics[width=\linewidth]{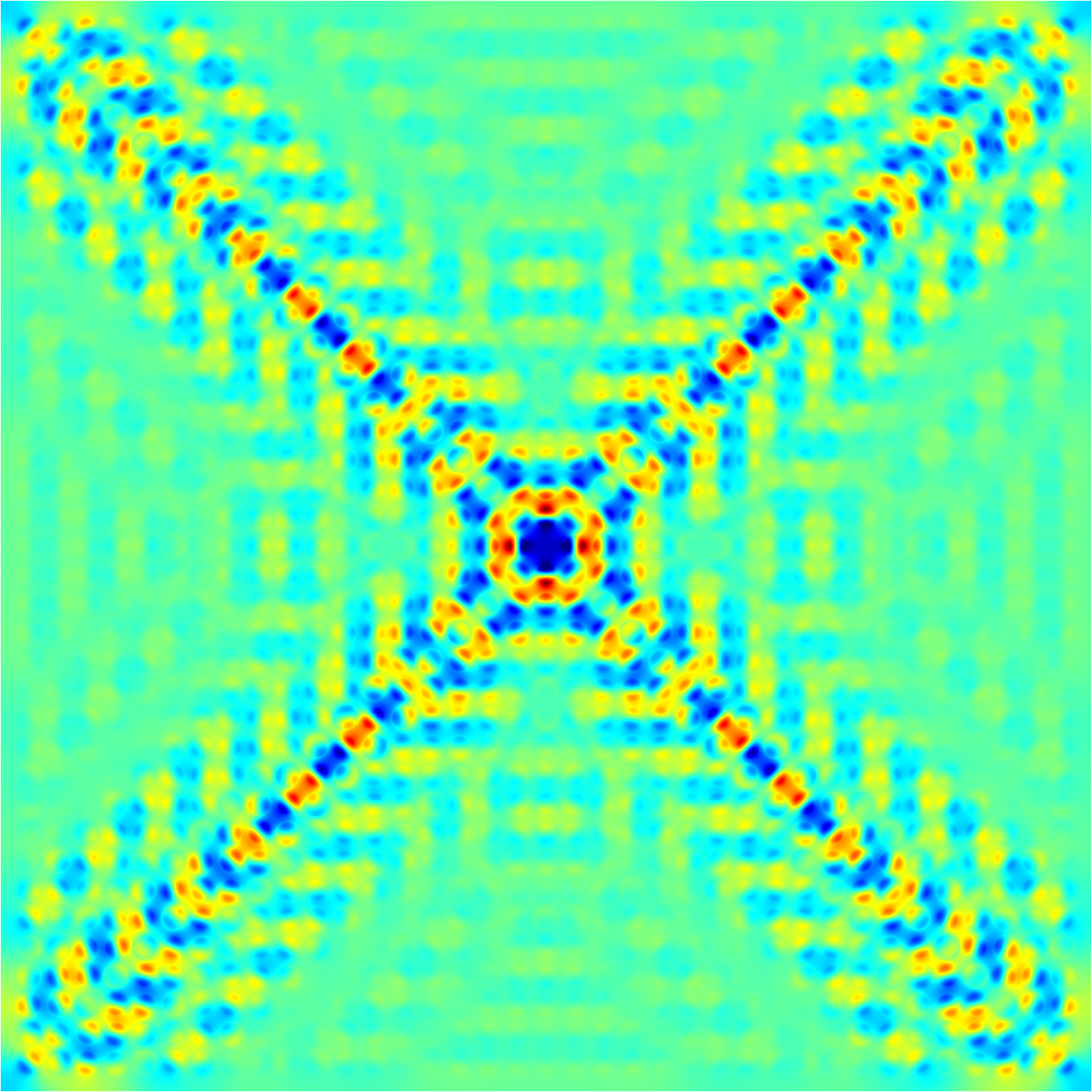}
\caption{\label{fig:RB-waves-numerics}\scriptsize
Finite element simulation}
\end{subfigure}
\begin{subfigure}{0.9\linewidth}
\includegraphics[width=\linewidth]{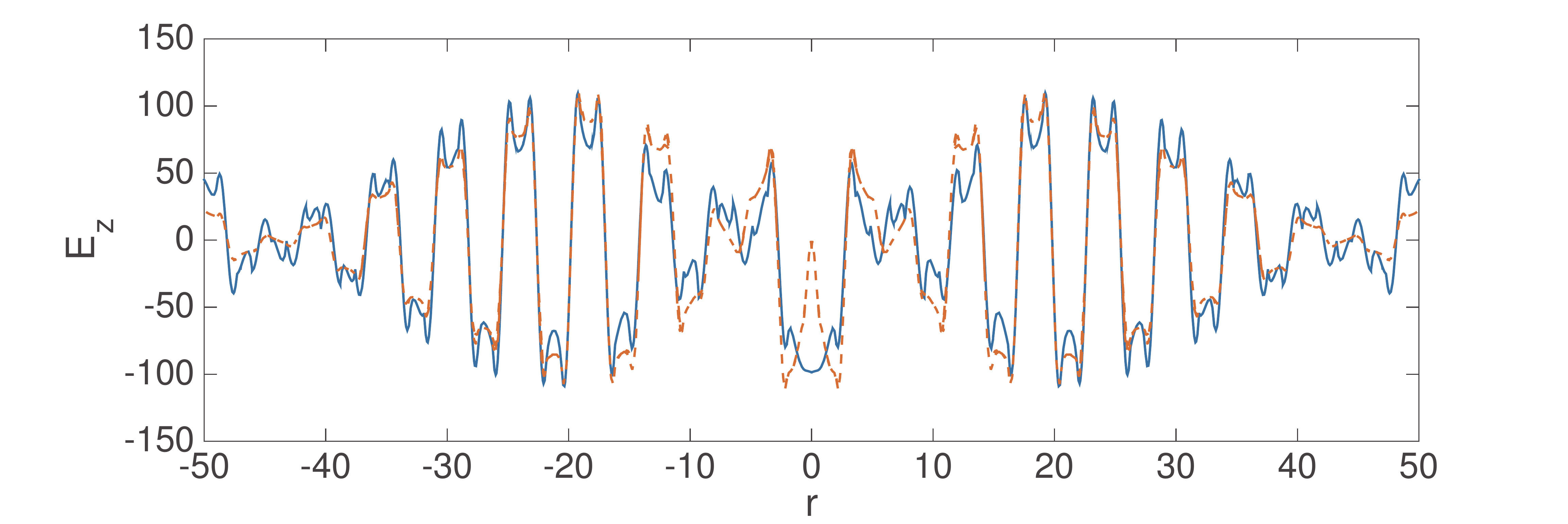}
\caption{\label{fig:RB-line-comp}}
\end{subfigure}
\caption{\label{fig:RB-comp}\scriptsize
The $z$-component of the electric field for the planar array obtained from (a) the asymptotic analysis and (b) the full finite element simulations.
In both cases the colour scale is linear running for minimum (blue), through zero (green) to maximal (red).
Part (c) shows the $z$-component of the electric field plotted along a line passing through the points $\boldsymbol{x} = (0,0,0)$ and $(1,1,0)$.
The solid curve represents the field from the full finite element simulation, whereas the dashed curve corresponds to the leading order solution from the asymptotic analysis.
The non-dimensional angular frequency of excitation is $\Omega = 0.80705$, which is close to the resonant frequency of $\Omega_0 = 0.8071$.}
\end{figure}

Using the asymptotic theory developed earlier, we first obtain the leading order solution for the field when the array is excited by a dipole source at a frequency close to that of the saddle point on the dispersion surface ($\Omega_0 =0.8071$).
This involves solving the leading and first order cell problems as outlined in section \ref{sec:proc}, and the solvability condition for the second-order problem then generates a PDE, of the same form as~\eqref{pde8}, for the long-scale field:
\begin{equation}
 T_{ij}\frac{\partial^2f_0}{\partial x_i \partial x_j}+\frac{(\omega^2-\omega_0^2)}{c^2}f_0=0,
\end{equation}
where $T$ is now a rank-2 tensor. In this case, we find that $T=\operatorname{diag}(-0.19,0.11)$ is diagonal and $T_{11}T_{22} < 0$, yielding a hyperbolic PDE on the long-scale.
The leading order asymptotic field for the forced problem is then a superposition of the solution to this PDE, whose forcing must take into account the symmetry of the cell problem, along with its symmetrical counterpart excited at point $Y$ in the Brillouin zone. The result is most striking when the electric field is plotted, as in figure~\ref{fig:RB-waves-HFH}, where since the hyperbolic mode is primarily out-of-plane electric, hence we only include the $E_3$-component of the field.
Here, the non-dimensional angular frequency of excitation is $\Omega = 0.80705$, which lies close to the resonant frequency of $\Omega_0 = 0.8071$.
Both the cell and long-scale problems were solved using the commercial finite element package Comsol Multiphysics\textsuperscript{\textregistered}.

For comparison the above problem is also treated, purely numerically, using finite elements.
In particular, a $37\times37$ planar array of spheres is considered; the finite cluster is surrounded by regions of perfectly matched layers in order to simulate an infinite domain.
Moreover, the computational cost of the full finite element simulation is lessened by making full use of the available symmetries, thereby reducing the computational domain by a factor of $8$.
The array is then excited, at the centre of the array, by a dipole source of unitary magnitude and oriented along the $x_3$-axis.
The $E_3$ component of the electric field from the full numerical simulation is shown in figure~\ref{fig:RB-waves-numerics} and should be compared with the asymptotic field shown in figure~\ref{fig:RB-waves-HFH}.
For comparison, we also plot the field along the line $(x,y,z) = (x,x,0)$ and passing through the dipole source (see figure~\ref{fig:RB-line-comp}).

\begin{figure}
\centering
\includegraphics[width=0.5\linewidth]{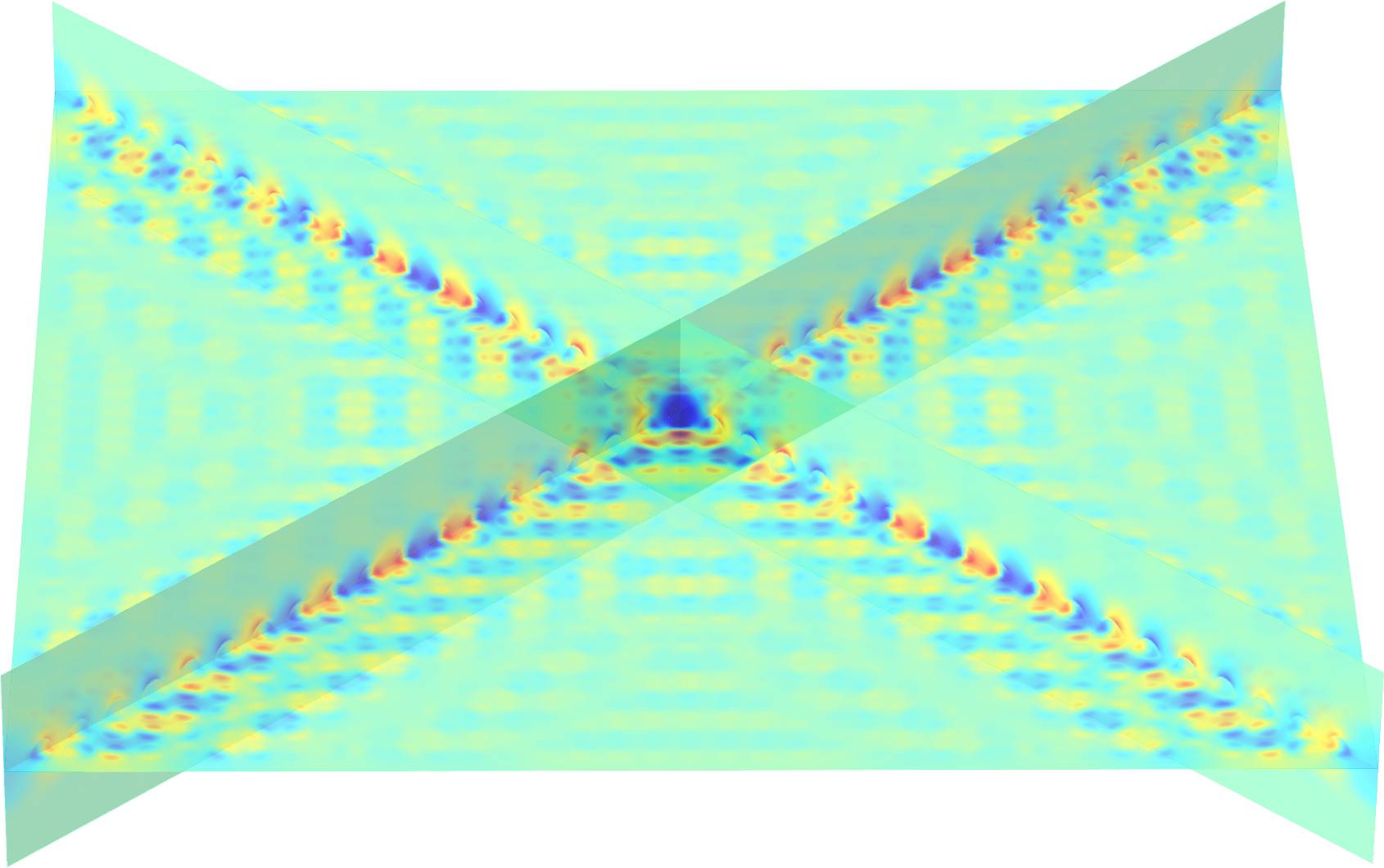}
\caption{\label{fig:RB-3D-slice}\scriptsize
Slices through the array showing the $E_3$ component of the electric field and illustrating the decay of the surface wave in the direction perpendicular to the array together with the in-plane dynamic anisotropy.
Once again, the colour scale is linear running for minimum (blue), through zero (green) to maximal (red).}
\end{figure}

Finally, figure~\ref{fig:RB-3D-slice} shows the $E_3$ component of the electric field both in the plane of the array and, over two slices, perpendicular to the array.
The figure illustrates the decay of the field in the direction perpendicular to the array in addition to the dynamic anisotropy exhibited in the plane of the array.
The novel exhibition of dynamic anisotropy on a structured metafilm is an exciting effect with several potential applications in the guiding of surface waves; this effect allows not only the confinement of waves to a surface or interface, but also the control of their propagation within the metafilm itself.
These effects occur, by necessity, at frequencies where the wavelength is comparable to the size of the microstructure where traditional (long-wave) homogenisation theories are no longer valid. 
However, as demonstrated by figure~\ref{fig:RB-comp}, the asymptotic homogenisation scheme developed here is capable of accurately and conveniently describing such effects. 

Although we have not carried out a comprehensive analysis of the computational cost of the asymptotic method, it is illuminating to consider the following: 
The full finite element simulation, using all the available symmetries, required approximately 16.8 million degrees of freedom and took 11.5 hours to solve on a dedicated machine using 77GB of RAM and 16 processor cores running at 2.5GHz.
In comparison, the numerical element of the asymptotic scheme used around 830 thousand degrees of freedom, 3.7GB of RAM and required 2.7 minutes to solve on a laptop with 2 cores running at 2.5GHz.

\section{Concluding remarks}
\label{sec:conclude}

We have demonstrated that in the vicinity of Brillouin zone vertices for a periodic structure, photonic modes governed by the vector Maxwell system give rise to second-order scalar PDEs, whose solutions govern the propagation, or decay, of waves inside the structure.  In some cases, these PDEs can be used straightforwardly to predict qualitatively and quantitatively the long-scale behaviour of a periodic photonic structure, which has been demonstrated in the context of a forced problem in a planar PC, as well as an eigenvalue problem for a localised mode in a PCF.

A particularly interesting physical example was considered in section \ref{sec:RB} wherein the asymptotic homogenisation theory was applied to a fully coupled three-dimensional problem for the Maxwell system.
Using a planar array of dielectric spheres we were able to create an effective hyperbolic metafilm which supports surface waves that also exhibit dynamic anisotropy over the surface.
One can envisage many applications where the ability to guide electromagnetic waves along a surface whilst also controlling the propagation over the surface itself would be useful.

A natural extension of this work is to consider more general Bravais geometries that are not necessarily square/cubic. In fact, such work has recently been completed for the simpler case of TE-polarised radiation\cite{mehul14}, and it is clear that the majority of the theory presented here carries through mostly unaltered. A second extension would be to consider structures with a wider range of material properties, for example those experiencing dielectric loss, or perhaps more interestingly real metals, whose frequency-dependent permittivity can be described by the Drude model.

\section*{Acknowledgements}
The authors thank the EPSRC for support
through research grants (EP/I018948/1, EP/L024926/1, EP/J009636/1) and Mathematics Platform
grant (EP/I019111/1).

\begin{appendices}



\section{The $\mathcal{O}(\eta)$ compatibility condition}\label{sec:Oe}

For each $n\in\{1,2,...p\}$, we first take the scalar product of (\ref{Oe}) with $\mathbf{h}_0^{(n)*}$ and subtract from the scalar product of (\ref{O1})$^*$ with $\mathbf{H}_1/f_0^{(n)}$. The resulting equation reads

\begin{equation}\begin{split}\label{compat2}
\mathbf{h}_0^{(n)*}\cdot\big(\nabla_{\xi}&\times\epsilon_{\text{r}}^{-1}\nabla_{\xi}\times \mathbf{H}_1\big)-\mathbf{H}_1\cdot\big(\nabla_{\xi}\times\epsilon_{\text{r}}^{-1}\nabla_{\xi}\times \mathbf{h}_0^{(n)*}\big)\\&=\mathbf{h}_0^{(n)*}\cdot\big(-\nabla_{\xi}\times\epsilon^{-1}\nabla_X\times \mathbf{H}_0-\epsilon_{\text{r}}^{-1}\nabla_{X}\times\nabla_\xi\times\mathbf{H}_0+\mu_{\text{r}}\Omega_1^2\mathbf{H}_0\big).
\end{split}\end{equation}

 Using the vector identity $\mathbf{A}\cdot(\nabla\times\nabla\times\mathbf{B})-\mathbf{B}\cdot(\nabla\times\nabla\times\mathbf{A}) =\nabla\cdot\big(\mathbf{B}\times\nabla\times\mathbf{A}-\mathbf{A}\times\nabla\times\mathbf{B}\big)$~\cite{vector}, the left hand side of this is 

\begin{equation}\label{compat4}
\epsilon_{\text{r}}^{-1}\nabla_\xi\cdot\big(\mathbf{H}_1\times\nabla_\xi\times\mathbf{h}_0^{(n)*}-\mathbf{h}_0^{(n)*}\times\nabla_\xi\times\mathbf{H}_1\big)
+(\nabla_\xi \epsilon_{\text{r}}^{-1})\cdot\big(\mathbf{H}_1\times\nabla_\xi\times\mathbf{h}_0^{(n)*}-\mathbf{h}_0^{(n)*}\times\nabla_\xi\times\mathbf{H}_1\big),
\end{equation}

 which is simply the divergence of the quantity $\epsilon_{\text{r}}^{-1}\big(\mathbf{H}_1\times\nabla_\xi\times\mathbf{h}_0^{(n)*}-\mathbf{h}_0^{(n)*}\times\nabla_\xi\times\mathbf{H}_1\big)$. Integrating this over the elementary cell $\mathcal{C}$ and applying the divergence theorem leaves us with an integral over the surface of the cell which vanishes by periodicity/antiperiodicity of $\mathbf{H}_1$ and $\mathbf{h}_0^{(n)}$, along with a non-vanishing contribution at the phase boundaries given by

\begin{equation}\label{compat5}
\int_{\cup\partial \mathcal{D}_{1,2}}\big[\epsilon_{\text{r}}^{-1}\big\{\mathbf{h}_0^{(n)*}\cdot\big(\mathbf{n}\times\nabla_\xi\times\mathbf{H}_1\big)-\mathbf{H}_1\cdot\big(\mathbf{n}\times\nabla_\xi\times\mathbf{h}_0^{(n)*}\big)\big\}\big]\mathrm{d}S,
\end{equation}

 where again square brackets denote a jump discontinuity. The boundary conditions in (\ref{BCO1}) and (\ref{BCOe}) ensure that the only non-zero contribution comes from the first term above, and we are left with the following equation:

\begin{equation}\label{compat6}\begin{split}
\int_{\cup\partial \mathcal{D}_{1,2}}\big[\epsilon_{\text{r}}^{-1}\mathbf{h}_0^{(n)*}&\cdot\big(\mathbf{n}\times\nabla_\xi\times\mathbf{H}_1\big)\big]\mathrm{d}S\\
&=\int_\mathcal{C}\mathbf{h}_0^{(n)*}\cdot\big\{-\nabla_{\xi}\times\epsilon_{\text{r}}^{-1}\nabla_X\times \mathbf{H}_0-\epsilon_{\text{r}}^{-1}\nabla_{X}\times\nabla_\xi\times\mathbf{H}_0+\mu_{\text{r}}\Omega_1^2\mathbf{H}_0\big\}\mathrm{d}V,
\end{split}\end{equation}

 for $n=1,2,...,p$. Using the cyclic property of the scalar triple product, the first term of the integrand on the right is  $\nabla_\xi\cdot\big(\mathbf{h}_0^{(n)*}\times\epsilon_{\text{r}}^{-1}\nabla_X\times\mathbf{H}_0\big)-\epsilon_{\text{r}}^{-1}\big(\nabla_\xi\times\mathbf{h}_0^{(n)*}\big)\cdot\big(\nabla_X\times\mathbf{H}_0\big)$. Applying the divergence theorem to the first resulting term yields a surface integral which, with the help of the boundary conditions in (\ref{BCO1}) and (\ref{BCOe}), exactly cancels the surface integral on the left hand side of (\ref{compat6}), leaving the equation

\begin{equation}\label{compat7}
0=\int_\mathcal{C}\big\{-\epsilon_{\text{r}}^{-1}\big(\nabla_\xi\times\mathbf{h}_0^{(n)*}\big)\cdot\big(\nabla_X\times\mathbf{H}_0\big)-\epsilon_{\text{r}}^{-1}\mathbf{h}_0^{(n)*}\cdot\nabla_{X}\times\nabla_\xi\times\mathbf{H}_0+\mu_{\text{r}}\Omega_1^2\mathbf{h}_0^{(n)*}\cdot\mathbf{H}_0\big\}\mathrm{d}V.
\end{equation}

 Now, substituting the solution $\mathbf{H}_0=f_0^{(r)}\mathbf{h}_0^{(r)}$ and noting that $\epsilon_{\text{r}}^{-1}\nabla_\xi\times\mathbf{h}_0^{(n)}=-i\Omega_0\mathbf{e_0}^{(n)}$ leads to the system of equations (\ref{Peq}).

\section{The $\mathcal{O}(\eta^2)$ compatibility condition: effective PDEs}\label{sec:Oee}

We now derive a compatibility condition from (\ref{Oee}) following a similar methodology to that of appendix \ref{sec:Oe}; we begin by taking the scalar product of (\ref{Oee}) with $\mathbf{h}_0^{(n)*}$, subtract from the scalar product of (\ref{O1})$^*$ with $\mathbf{H}_2/f_0^{(n)}$, and integrate the resulting equation over the elementary cell $\mathcal{C}$. The left hand side vanishes like the first order equivalent, leaving 

\begin{equation}\begin{split}\label{pde1}
0=\int_{\mathcal{C}}\big\{-\epsilon_{\text{r}}^{-1}\big(\nabla_\xi\times\mathbf{h}_0^{(n)*}\big)&\cdot\big(\nabla_X\times\mathbf{H}_1\big)-\epsilon_{\text{r}}^{-1}\mathbf{h}_0^{(n)*}\cdot\nabla_{X}\times\nabla_\xi\times\mathbf{H}_1\\&-\epsilon_{\text{r}}^{-1}\mathbf{h}_0^{(n)*}\cdot\nabla_{X}\times\nabla_X\times \mathbf{H}_0+\mu_{\text{r}}\Omega_2^2\mathbf{h}_0^{(n)*}\cdot\mathbf{H}_0\big\}\mathrm{d}V.
\end{split}\end{equation}

 Using the cyclic property of the scalar triple product, it is straightforward to show that the contribution of the complementary part of $\mathbf{H}_1$ in the above equation is zero. Changing to tensor notation, and making use of the contracted epsilon identity $\varepsilon_{ijk}\varepsilon_{klm}=\delta_{il}\delta_{jm}-\delta_{im}\delta_{jl}$, we are led to the following system of effective PDEs:

\begin{equation}\label{sys8}
\hat{T}_{ij}^{nr}\frac{\partial^2f_0^{(r)}}{\partial X_i \partial X_j}+\Omega_2^2Q^{nr}f_0^{(r)}=0\Longleftrightarrow T_{ij}^{nr}\frac{\partial^2f_0^{(r)}}{\partial X_i \partial X_j}+\Omega_2^2f_0^{(n)}=0,
\end{equation}

 where $T_{ij}^{nr}=(Q^{-1})^{nq}\hat{T}_{ij}^{qr}$, with $Q^{nr}=\int_{\mathcal{C}} \mu_{\text{r}}h_{0k}^{(n)*}h_{0k}^{(r)}\mathrm{d}V$ and

\begin{equation}\label{pde9}\begin{split}
\hat{T}_{ij}^{nr}=\int_{\mathcal{C}}\epsilon_{\text{r}}^{-1}\Bigg\{&\delta_{ij}h_{0k}^{(n)*}h_{0k}^{(r)}-h_{0j}^{(n)*}h_{0i}^{(r)}\\&+h_{1kj}^{(r)}\partial_kh_{0i}^{(n)*}-h_{1kj}^{(r)}\partial_ih_{0k}^{(n)*}+h_{0k}^{(n)*}\partial_ih_{1kj}^{(r)}-h_{0k}^{(n)*}\partial_kh_{1ij}^{(r)}\Bigg\}\mathrm{d}V.
\end{split}\end{equation}

In the case of a distinct eigenvalue, the above system reduces to a single equation with $n=r=1$. For a repeated eigenvalue, the substitution $f_0^{(r)}(\mathbf{X})=\hat{f}_0^{(r)}\exp(i\bf{\kappa}\cdot\mathbf{X}/\eta)$ yields a homogeneous matrix equation whose solution gives the quadratic asymptotics, but in fact the independent Bloch modes are non-interacting in the sense that they can be decoupled as:

\begin{equation}\label{MTtensor}
M^{np}\tilde{T}_{ij}^{pq}(M^{-1})^{qr}\frac{\partial^2f_0^{(r)}}{\partial X_i \partial X_j}+\Omega_2^2f_0^{(n)}=0,
\end{equation}

 where $M$ is matrix of eigenvectors shared by the the matrices $T_{ij}$, and hence $\tilde{T}_{ij}$ are all diagonal matrices. Pre-multiplying by $M^{-1}$, we are then left with the decoupled system

\begin{equation}\label{Diag}
\tilde{T}_{ij}^{nr}\frac{\partial^2\tilde{f}_0^{(r)}}{\partial X_i \partial X_j}+\Omega_2^2\tilde{f}_0^{(n)}=0,
\end{equation}

 where $\tilde{f}_0^{(n)}=(M^{-1})^{nr}f_0^{(r)}$. We can write the general leading order solution as a linear combination of these decoupled modes $\mathbf{H}_0=\tilde{f}_0^{(r)}\tilde{\mathbf{h}}_\mathbf{0}^{(r)}$, with $\tilde{\mathbf{h}}_\mathbf{0}^{(n)}=M^{rn}\mathbf{h}_0^{(r)}$. Note that the tildes have been omitted in equation (\ref{pde8}) as it is assumed that the decoupling is done automatically.


\end{appendices}

\bibliographystyle{plain}

\bibliography{maling_2014_6}

\end{document}